\documentclass[useAMS,usenatbib]{mn2e} 
 
\usepackage{graphicx} 
\usepackage{txfonts} 
\usepackage{natbib} 
\newcommand\aj{AJ} 
\newcommand\apj{ApJ}

\newcommand\aap{A\&A} 
\newcommand\mnras{MNRAS} 
\newcommand\apjl{ApJ}     
\newcommand\pasp{PASP} 
\newcommand\nat{Nature} 
 
\newcommand\aapr{A\&AR} 
\newcommand\araa{AA\&A}

\title[]{Multiple stellar populations in Magellanic Cloud clusters. III. The first evidence of an extended main sequence turn-off 
in a young cluster: NGC\,1856}  
\author[A.\,P.\, Milone et al.] 
{A.\,P.\,Milone$^{1}$,
 L.\,R.\,Bedin$^{2}$,
 G.\,Piotto$^{2,3}$,
 A.\,F.\,Marino$^{1}$,
 S.\,Cassisi$^{4}$,
 A.\,Bellini$^{5}$,
\newauthor
 H.\,Jerjen$^{1}$,
 A.\,Pietrinferni$^{4}$,
 A.\,Aparicio$^{6,7}$,
 R.\,M.\,Rich$^{8}$
\\ 
$^{1}$Research School of Astronomy \& Astrophysics, Australian National University, Mt Stromlo Observatory, via Cotter Rd, Weston, ACT 2611, Australia \\
$^{2}$Istituto Nazionale di Astrofisica - Osservatorio Astronomico di Padova, Vicolo dell'Osservatorio 5, Padova, IT-35122\\
$^{3}$Dipartimento di Fisica e Astronomia ``Galileo Galilei'', Univ. di Padova, Vicolo dell'Osservatorio 3, Padova, IT-35122\\
$^{4}$Istituto Nazionale di Astrofisica - Osservatorio Astronomico di Teramo, Via Mentore  Maggini s.n.c., I-64100 Teramo, Italy\\
$^{5}$Space Telescope Science Institute, 3800 San Martin Drive, Baltimore,  MD 21218, USA\\
$^{6}$Instituto de Astrof\`\i sica de Canarias, E-38200 La Laguna, Tenerife, Canary Islands, Spain\\
$^{7}$Department of Astrophysics, University of La Laguna, E-38200 La Laguna, Tenerife, Canary Islands, Spain\\
$^{8}$Department of Physics and Astronomy, University of California, Los Angeles, CA 90095, USA\\
} 
\begin{document} 
%\date{Accepted xxx December 15. Received xxx December 14; in original form xx October 11} 
\date{Draft Version Apr, 3, 2015} 
 
\pagerange{\pageref{firstpage}--\pageref{lastpage}} \pubyear{2015} 
 
\maketitle 
\label{firstpage} 
 
\begin{abstract}  
 Recent studies have shown that the extended main-sequence turn off (eMSTO) is a common feature of intermediate-age star clusters in the Magellanic Clouds (MCs). The most simple explanation is that these stellar systems harbor multiple generations of stars with an age difference of a few hundred Myrs. 
 
 However, while an eMSTO has been detected in a large number of clusters with ages between $\sim$1-2 Gyrs, several studies of young clusters in both MCs and in nearby galaxies do not find any evidence for a prolonged star-formation history, i.\,e.\,for multiple stellar generations. 
These results have suggested alternative interpretation of the eMSTOs observed in intermediate-age star clusters. The eMSTO could be due to stellar rotation mimicking an age spread or to interacting binaries. In these scenarios, intermediate-age MC clusters would be simple stellar populations, in close analogy with younger clusters.  
 
Here we provide the first evidence for an eMSTO in a young stellar cluster. We exploit multi-band {\it Hubble Space Telescope} photometry to study the  $\sim$300-Myr old star cluster NGC\,1856 in the Large Magellanic Cloud and detected a broadened MSTO that is consistent with a prolonged star-formation which had a duration of about 150 Myrs. 

Below the turn-off, the MS of NGC\,1856 is split into a red and blue component, hosting 33$\pm$5\% and 67$\pm$5\% of the total number of MS stars, respectively.   
We discuss these findings in the context of multiple-stellar-generation, stellar-rotation, and interacting-binary hypotheses. 
\end{abstract} 
 
\begin{keywords} 
globular clusters: individual (NGC\,1856) --- stars: Population~II 
\end{keywords} 
 
\section{Introduction}\label{sec:intro} 
 The finding that intermediate-age star clusters in the Magellanic Clouds exhibit a bimodal or extended main-sequence turn off (eMSTO) is one of the most intriguing discoveries in the field of stellar populations of the last decade (e.g.\,Bertelli et al.\,2003; Baume et al.\,2007; Mackey \& Broby Nielsen\,2007; Glatt et al.\,2008; Goudfrooij et al.\,2011; Keller et al.\,2012; Rubele et al.\,2013).   
 High-accuracy photometry with the {\it Hubble Space Telescope} ({\it HST}) has revealed that the eMSTO is a common feature among the $\sim$1-2 Gyr-old MCs' clusters as it has been detected in most of the objects studied so far (Milone et al.\,2009 --- hereafter Paper\,I ---; Goudfrooij et al.\,2014). 
 
 The most straightforward interpretation of the eMSTO is that these stellar systems have experienced an extended star-formation history with a duration of $\sim$100-500 Myr (e.g.\,Mackey et al.\,2008; Paper\,I; Goudfrooij et al.\,2009, 2014). As an alternative explanation, Bastian \& de Mink\,(2009) and Yang et al.\,(2011, 2013) have suggested that stellar rotation or interacting binaries can mimic an age spread and could be responsible for the eMSTO (but see Girardi et al.\,2011  and Platais et al.\,2012). 
 
Noticeably, there is no evidence for prolonged star formation in clusters younger than $\sim$1 Gyr although age spreads have been searched in a large number of young massive clusters with different techniques (Larsen et al.\,2011; Milone et al.\,2013 --- hereafter Paper\,II ---; Bastian \& Silva-Villa\,2013; Bastian et al.\,2013; Niederhofer et al.\,2015). 
 
The properties of the progenitors of the clusters where we observe eMSTO have been widely studied by Bekki \& Mackey\,(2009) and Keller et al.\,(2011) in the context of a prolonged star-formation history.    
 A sufficiently large mass should enable a cluster, in principle, to retain ejecta from a previous stellar generation which polluted the gas
from which the second generation can form. This possibility implies that the mass of the progenitor should have been of the order of $10^{6} \mathcal{M}_{\odot}$ at the time of cluster formation.  
 Keller et al.\,(2011) concluded that NGC\,1856 in the Large Magellanic Cloud, with a mass of $\sim 10^{5} \mathcal{M}_{\odot}$ (Mackey \& Gilmore\,2003) and an age of $\sim$300 Myr (Bastian \& Silva-Villa\,2013) is the only cluster with measured age and mass close to that of an eMSTO progenitor. 
 
 NGC\,1856 has been previously studied by Bastian \& Silva-Villa\,(2013) who have analyzed Wide-Field-Planetary Camera 2 (WFPC2) photometry  from Brocato et al.\,(2001) and found an age of 280 Myr and
no evidence for an eMSTO and an age spread. 
 In this paper we exploit a new dataset, which consists of multi-wavelength images collected with the Ultraviolet and Visual channel of the Wide Field Camera 3 (UVIS/WFC3) on board of {\it HST}, to further investigate NGC\,1856 in the context of multiple stellar populations.  
 
\section{Data and data analysis} 
\label{sec:data} 
The dataset used in this paper is summarized in Table~\ref{tab:data} and consists of images collected with UVIS/WFC3 through five filters that cover a wavelength range from the ultraviolet to $\sim$8100 \AA.  
The poor charge-transfer efficiency (CTE) in the UVIS/WFC3 images have been corrected by using a software written by Jay Anderson following the recipe by Anderson \& Bedin\,(2010). 
 
Photometry and astrometry of the images have been carried out with a software program mostly based on the Anderson et al.\,(2008) software adapted to UVIS/WFC3 by Jay Anderson. Briefly, we used two distinct methods to infer luminosity and position of bright and faint stars. Astrometry and photometry of bright stars have been performed independently in each image by using the best point-spread function (PSF) model available, and later combined. For that purpose, we used library PSF by Anderson et al.\,(in preparation) and accounted for small focus variations due to the `breathing' of {\it HST} by deriving a spatially-constant perturbation for each exposure.  
The flux and the position of each very faint stars can be determined more robustly by simultaneously fitting all the pixels in all the exposures. Moreover, we adopted the solution of Bellini \& Bedin (2009) and Bellini, Anderson \& Bedin\,(2011) to correct stellar positions for geometrical distortion. 
The photometry has been calibrated as in Bedin et al.\,(2005) and by using the zero points provided by the STScI web page for WFC3/UVIS\footnote{http://www.stsci.edu/hst/acs/analysis/zeropoints/zpt.py}. 
 
Since the study of multiple populations in star clusters requires high-precision photometry, we have used a number of indexes provided by the adopted software as diagnostics of photometry quality (see Anderson et al.\,2008 for details). Similarly with what was done in Papers\,I (see their Sect.\,2.1) and II, we have limited our study of NGC\,1856 to a sub-sample of stars that have small astrometric errors, are relatively isolated, and well fitted by the PSF.  
 
\begin{centering} 
\begin{figure*} 
 \includegraphics[width=11.2cm]{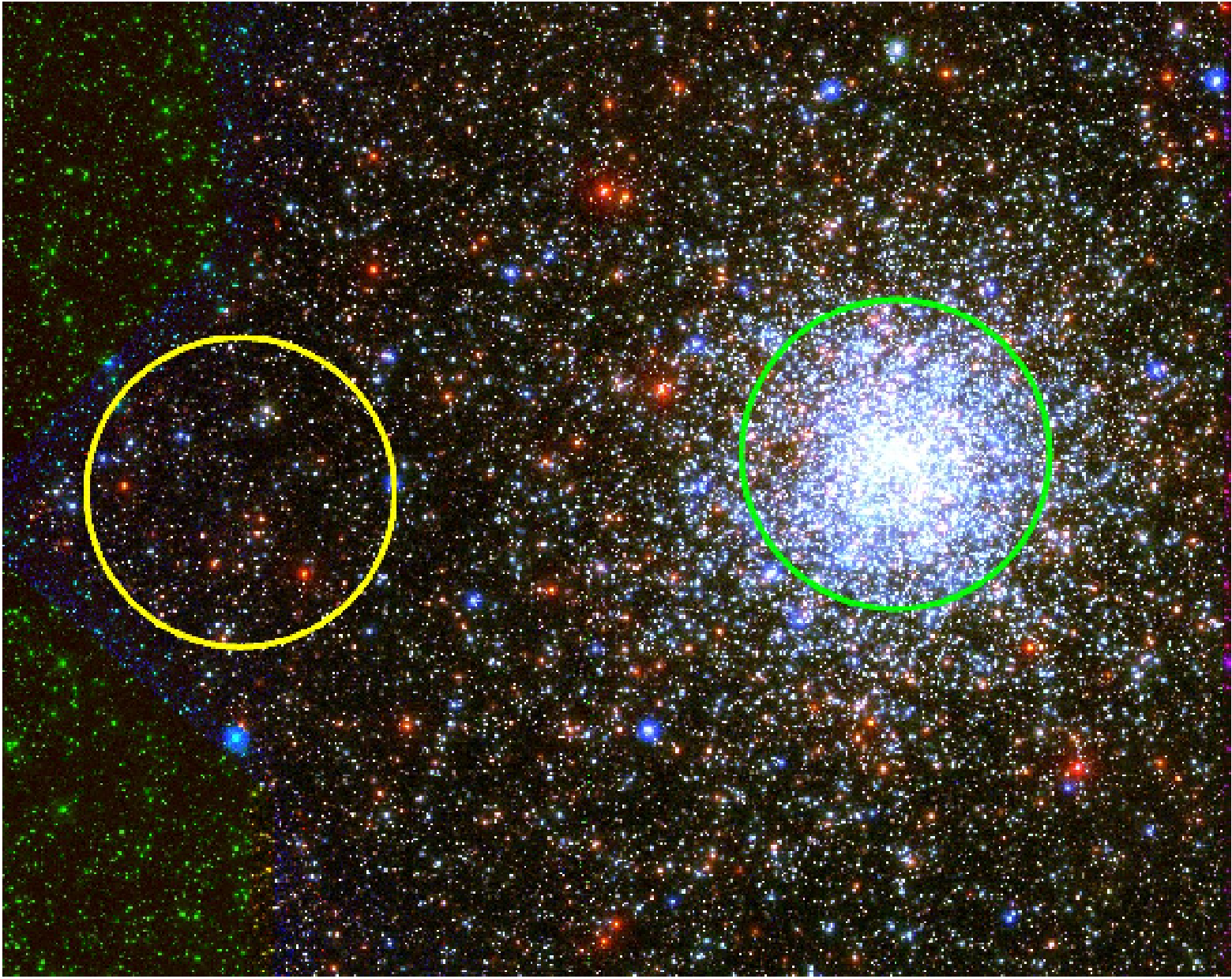} 
 \includegraphics[width=4.6cm]{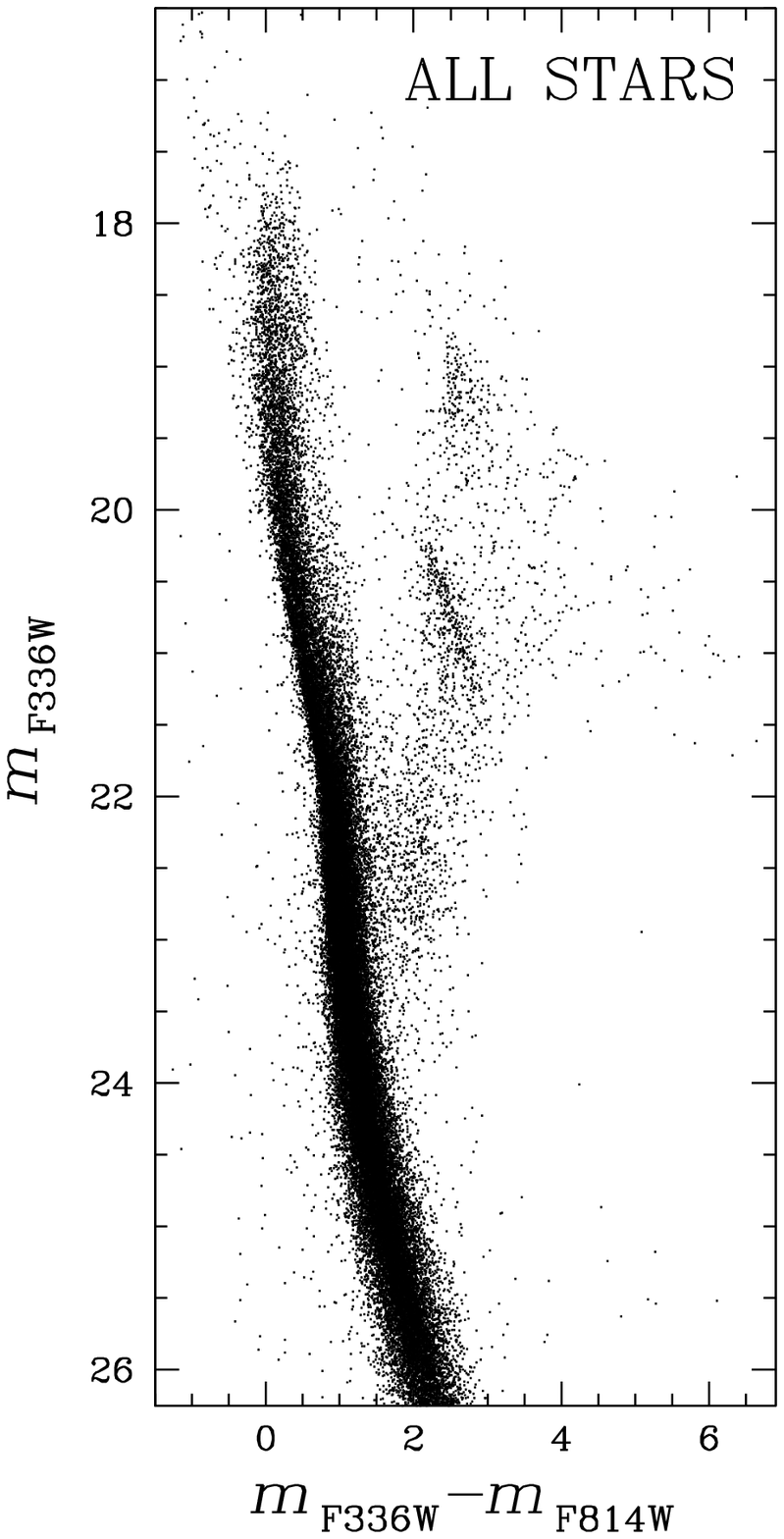} 
%/home/milone/MC/NGC1856/all/MATCH/figure/fig.macro go0
%/home/milone/MC/NGC1856/all/XKS/KSUV/macro/selezioni/cmd 
 \caption{\textit{Left panel:} Stacked trichromatic image of the field around NGC\,1856 studied in this paper. Green and yellow circles mark
the cluster and the reference field, respectively.      
          \textit{Right panel:} $m_{\rm F336W}$ vs.\,$m_{\rm F336W}-m_{\rm F814W}$ CMD of all the stars in the analyzed field of view for which F336W and F814W photometry is available.} 
 \label{fig:foot} 
\end{figure*} 
\end{centering}

 We show in Fig.~\ref{fig:foot} the trichromatic stacked images of the analyzed field of view (left panel) and the $m_{\rm F336W}$ vs.\,$m_{\rm F336W}-m_{\rm F814W}$ CMD of all the stars for which photometry in both F336W and F814W bands
 is available (right panel). 
 The two circles superimposed on the left-panel image have a radius of 25 arcsec and delimit a region named cluster field which is centered on NGC\,1856 and is mostly populated by cluster members (green circle), and a reference field, where the contamination from NGC\,1856 stars is negligible (yellow circle). Stars in these two regions will be used in the study of the multiple stellar populations of NGC\,1856 in Sects.~\ref{sec:cmd}-~\ref{sec:teoria}. 
 
%%%%%%%%%%%%%%%%%%%%%%%%%%%%%%%%%%%%%%%%%%%%%%%%%%%%%%%%%%%%%%%%%%%%%%%%%%%%%%% 
 
\subsection{Artificial Stars} 
We have used the method described in details
by Anderson et al.\,(2008) to perform the artificial-star (AS) tests.  
 Briefly, a list of 10$^{5}$ stars has been generated and placed along the fiducial line of the MS of NGC\,1856. The list includes the coordinates of the stars in the reference frame and the magnitudes in all the bands studied in this paper. ASs have been distributed across the field  of view according to the overall cluster distribution as in Paper\,I. 

In each image and for each star in the input list, the software by Anderson et al.\,(2008) generates a star with appropriate flux and position and measures it by using the same procedure as for real stars.  
 
For the ASs the software provides the same diagnostics of the photometric quality as for real star. We applied to ASs the same procedure as used for real stars to select a sub-sample of relatively-isolated stars with small astrometric errors, and well fitted by the PSF. 
ASs have been used to estimate errors of the photometry and estimate the fraction of chance-superposition binaries. Moreover, ASs have been used to derive the completeness level of our sample as in Paper\,I. 
 
\begin{table*} 
\caption{Description of the data set used in this paper.} 
\begin{tabular}{cccccc}%[ht!]                                                                                     
\hline\hline 
INSTR. & DATE & N$\times$EXPTIME & FILTER & PROGRAM & PI \\ 
\hline 
WFC3/UVIS & Feb 09, Mar 24, May 18, and Jun 11, 2014 & 11$\times$711s & F336W & 13379 & A.\,P.\,Milone \\ 
WFC3/UVIS & Nov 12 2013 & 185s$+$2$\times$430s & F438W & 13011 & T.\,H.\,Puzia \\ 
WFC3/UVIS & Nov 12 2013 & 2$\times$350s & F555W & 13011 &  T.\,H.\,Puzia\\ 
WFC3/UVIS & Nov 12 2013 & 735s$+$2$\times$940s & F656N & 13011 &  T.\,H.\,Puzia\\ 
WFC3/UVIS & Nov 12 2013 & 51s$+$360s$+$450s & F814W & 13011 &  T.\,H.\,Puzia\\ 
WFC3/UVIS & Feb 09, Mar 24, May 18, and Jun 11, 2014 & 4$\times$90s$+$4$\times$704s & F814W & 13379 & A.\,P.\,Milone \\ 
 
\hline\hline 
\end{tabular}\\ 
\label{tab:data} 
\end{table*} 
%%%%%%%%%%%%%%%%%%%%%%%%%%%%%%%%%%%%%%%%%%%%%%%%%%%%%%%%%%%%%%%%%%%%%%%%%%       
 
\section{The CMD of NGC\,1856}\label{sec:cmd} 
 
%%%%%%%%%%%%%%%%%%%%%%%%%%%%%%%%%%%%%% FIG 1 %%%%%%%%%%%%%%%%%%%%%%%%%%%%%%%%%%% 
\begin{centering} 
\begin{figure*} 
 \includegraphics[width=14.5cm]{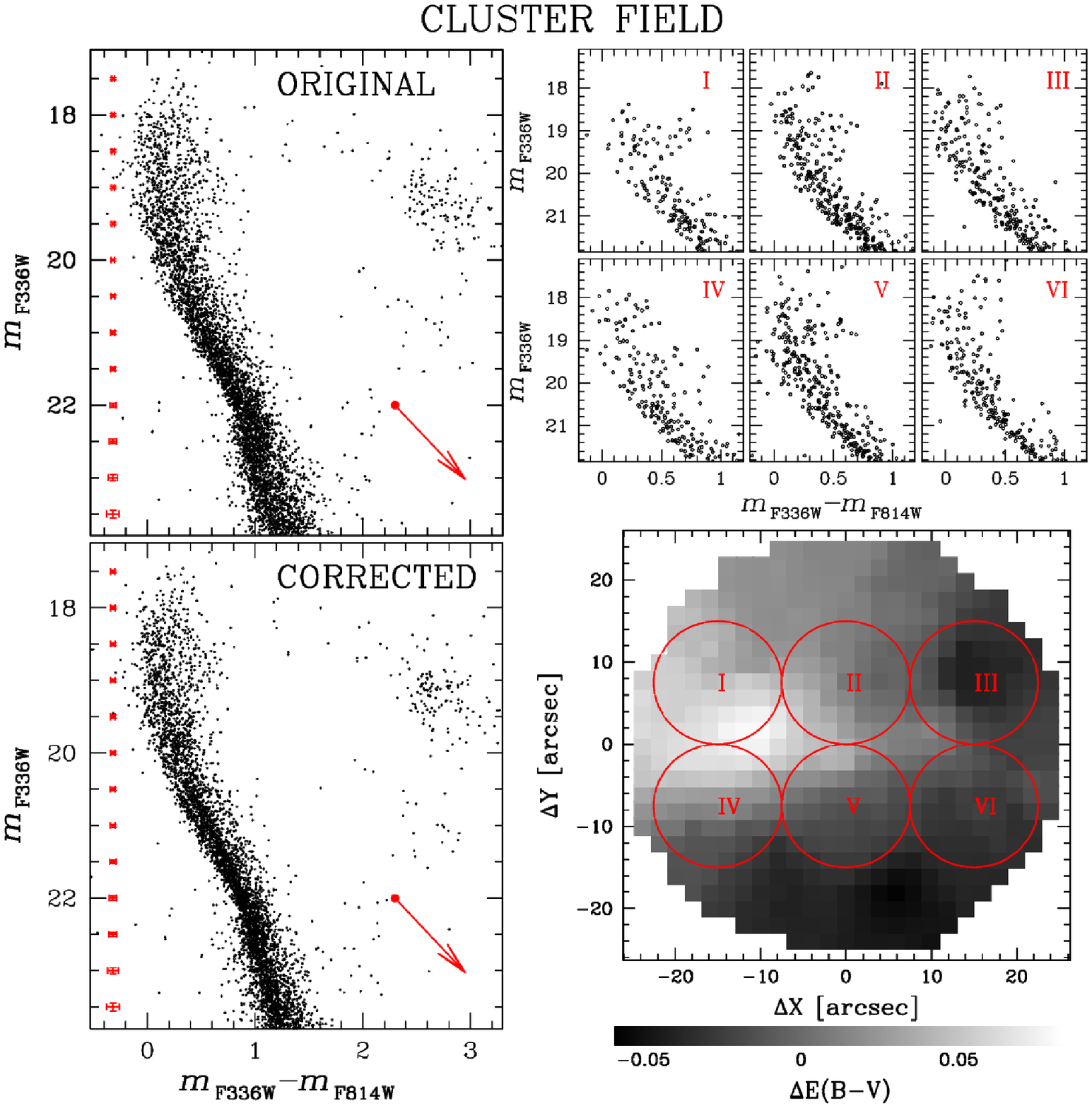} 
%/home/milone/MC/NGC1856/all/MATCH/figure/fig.macro go 
 \caption{\textit{Left panels:} Original $m_{\rm F336W}$ vs.\,$m_{\rm F336W}-m_{\rm F814W}$ CMD of stars within the NGC\,1856 cluster field (green circle in Fig.~\ref{fig:foot}, up), and CMD corrected for differential reddening (bottom). The red arrows indicate the reddening direction.  \textit{Right panels:} Original CMD of stars in six sub-regions, I-VI, within the cluster field (up). Map of differential reddening for the cluster field. The levels of  gray 
indicate the differential reddening as in the scale plotted
on the bottom of the figure. The six sub-regions are highlighted
with red circles.} 
 \label{fig:reddening} 
\end{figure*} 
\end{centering} 
%add error bars 
%%%%%%%%%%%%%%%%%%%%%%%%%%%%%%%%%%%%%%%%%%%%%%%%%%%%%%%%%%%%%%%%%%%%%%%%%%%%%%% 
 
The $m_{\rm F336W}$ vs.\,$m_{\rm F336W}-m_{\rm F814W}$ CMD of stars in the cluster field is plotted in the upper-left panel of Fig.~\ref{fig:reddening}. It shows that the MS and the MSTO of NGC\,1856 are widely spread in color and magnitude. 
 A visual comparison between the observed CMD and the error bars, derived from AS tests and plotted on the left, reveals that this spread can not entirely be explained by the photometric errors.
 In the following we investigate whether this broadening is intrinsic or can be due to differential reddening, spatial variation of the photometric zero point across the field of view, field stars, or binaries. 
 
\subsection{ Differential reddening } 
In order to investigate the effect of differential reddening on the MSTO of NGC\,1856, in the upper-right panels of Fig.~\ref{fig:reddening} we have plotted the CMD not corrected for differential reddening for the stars located
in six circular sub-regions (I-VI) of 25-arcsec radius.
Due to the small area of each region, the variation of reddening therein is smaller than that of the entire cluster field. 
 Noticeably the color and magnitude of the MSTO is broadened in each CMD, thus suggesting that the young NGC\,1856 hosts an eMSTO in close analogy with intermediate-age MC clusters. 
 
To properly address this issue, we have corrected the photometry of stars in the cluster field for differential reddening as in Milone et al.\,(2012).  
 Briefly, we started to select a sample of reference stars including those bright MS stars which pass the criteria of selection described in Sect.~\ref{sec:data}. Then we have determined the fiducial line of these reference stars and calculated, for each of them,
the distance from the fiducial line along the reddening line.  
The reddening direction has been determined by using the absorption coefficient for UVIS/WFC3 filters kindly provided by Aaron Dotter (2014, private communication). Namely these are $A_{\rm F336W}/E(B-V)=5.10$, $A_{\rm F438W}/E(B-V)=4.18$, $A_{\rm F555W}/E(B-V)=3.27$, $A_{\rm F656N}/E(B-V)=2.54$, and $A_{\rm F814W}/E(B-V)=1.86$. 
As differential reddening of each star in the cluster field we assumed  the median distance of the 45 closest reference stars. The uncertainty on the differential reddening has been estimated as the ratio between the dispersion of the 45 values of the
distance of each reference star from the fiducial line 
and the square root of 44. 
 We have found that the reddening in the direction of NGC\,1856 is not uniform and ranges from $\Delta$$E$($B-V$)$\sim$$-$0.05 to $\sim$$+$0.07 with respect to the average reddening $E$($B-V$)=0.15 (see Sect.~\ref{subsec:age}). 
 
The $m_{\rm F336W}$ vs.\,$m_{\rm F336W}-m_{\rm F814W}$ CMD corrected for differential reddening is shown in the lower-left panel of Fig.~\ref{fig:reddening}. The error bars plotted on the left account also for the uncertainty in the differential reddening correction above estimated. 
The corrected photometry unveils several features in the CMD of NGC\,1856.   
 
The color width of the MS, below $m_{\rm F336W}\sim$21.5, is comparable to the measurement errors and a large number of binary systems is visible on the red side of the MS. The MSTO is widely spread in color and magnitude thus indicating that NGC\,1856 belongs to the class of clusters with an eMSTO. 
 Furthermore, we confirm previous findings by Bastian et al.\,(2013) that the red clump of NGC\,1856 spans a wide color range and hosts two main components clustered around ($m_{\rm F336W}-m_{\rm F814W}$:$m_{\rm F336W}$)=(2.60:19.25) and (2.75:19.25). 
 
For completeness, we show the reddening map  in the lower-right panel of Fig.~\ref{fig:reddening}, where we also mark the sub-regions I-VI, used to derive the six CMDs in the upper-right panel. 
      
%%%%%%%%%%%%%%%%%%%%%%%%%%%%%%%%%%%%%% FIG 2 %%%%%%%%%%%%%%%%%%%%%%%%%%%%%%%%%%% 
\begin{centering} 
\begin{figure*} 
 \includegraphics[width=8.5cm]{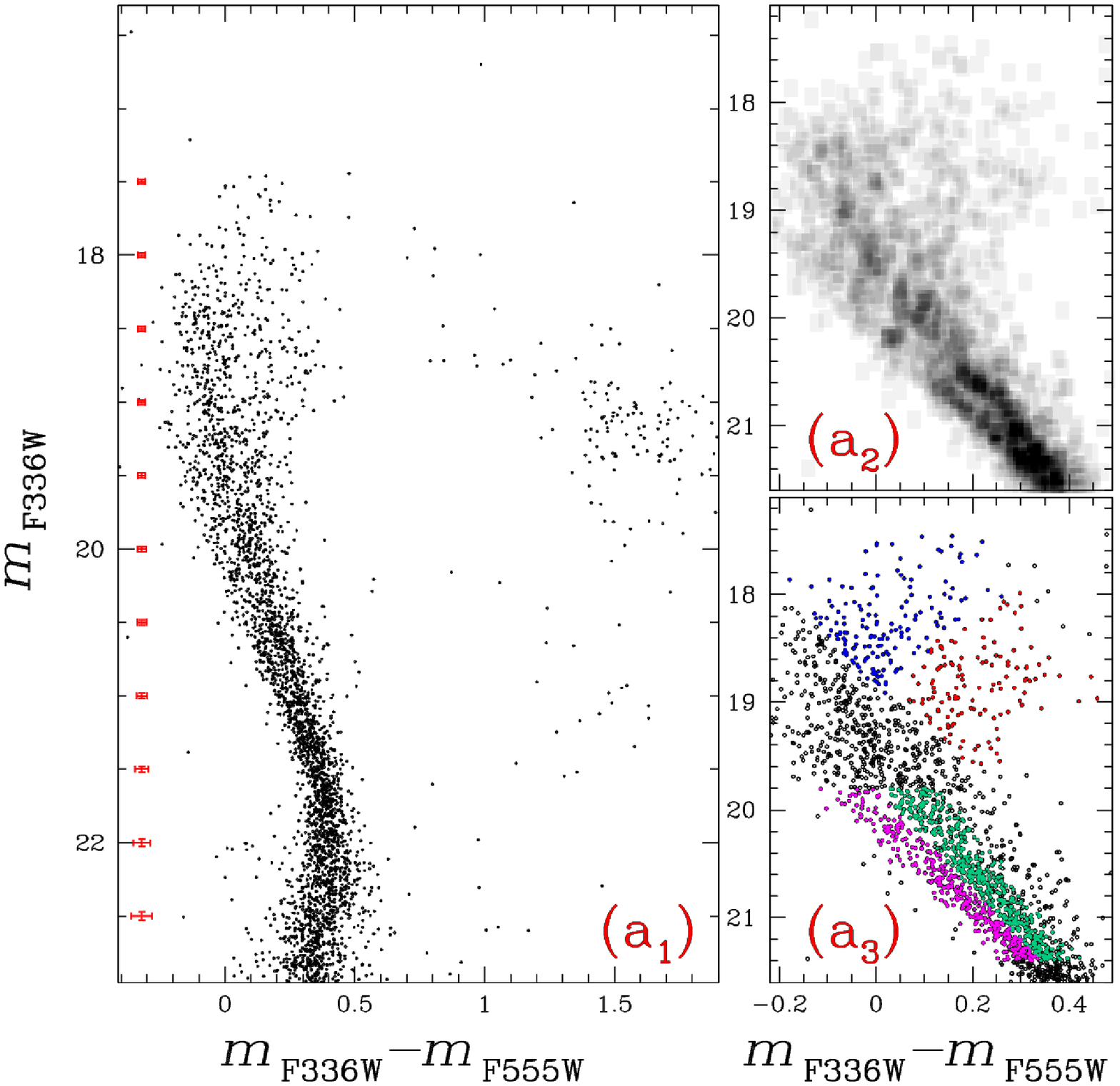} 
 \includegraphics[width=8.5cm]{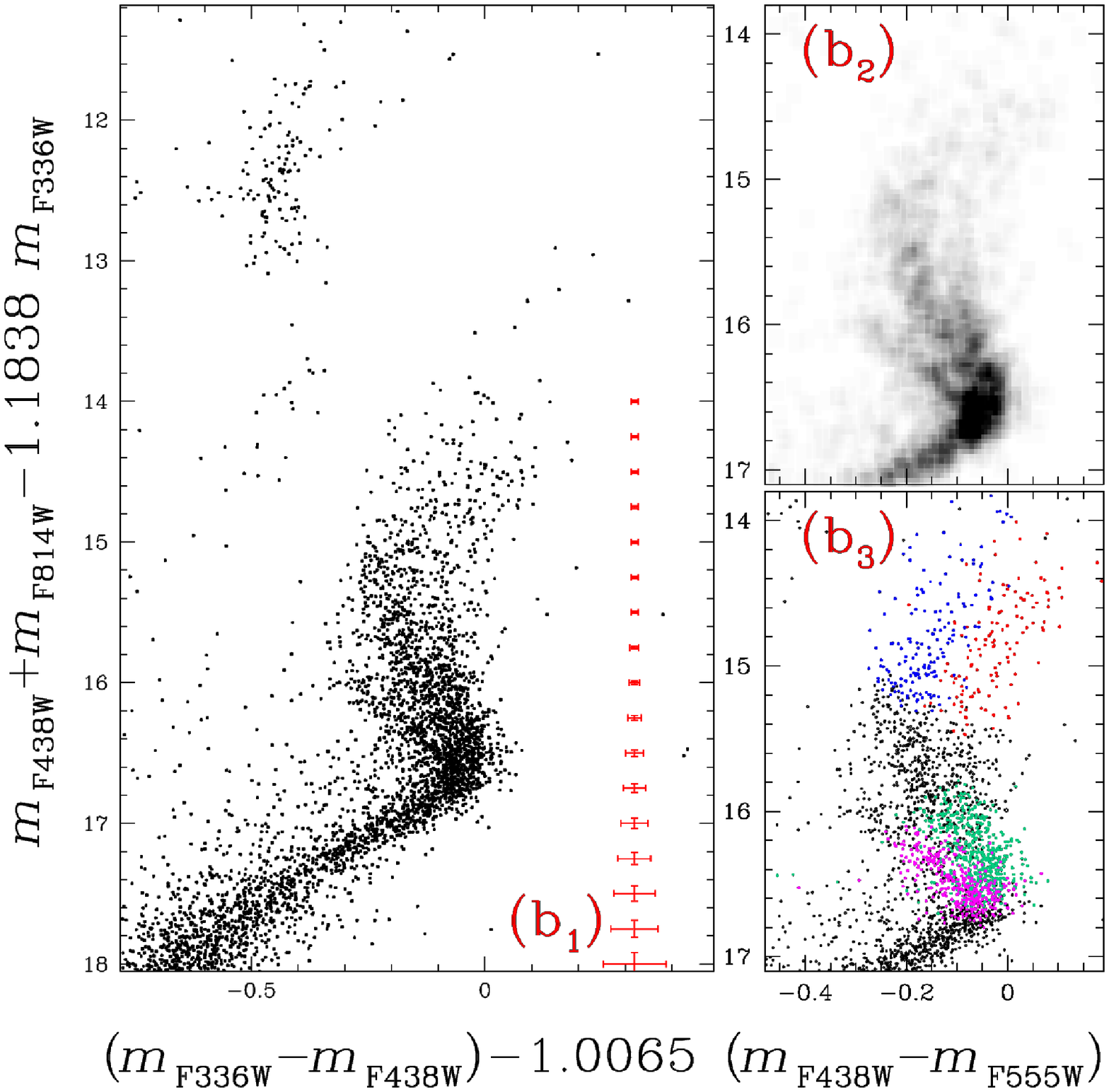} 
 \includegraphics[width=8.5cm]{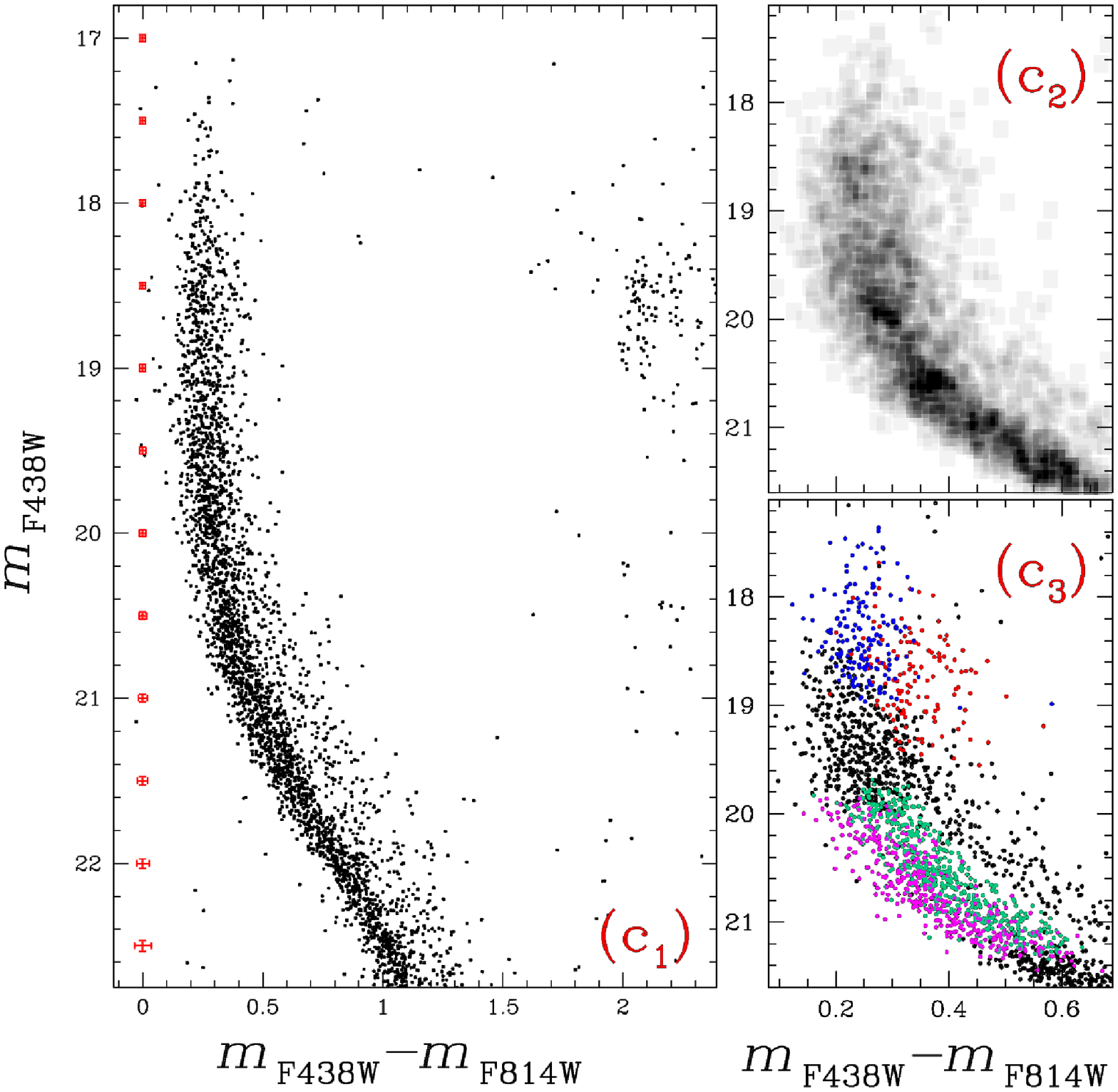} 
 \includegraphics[width=8.5cm]{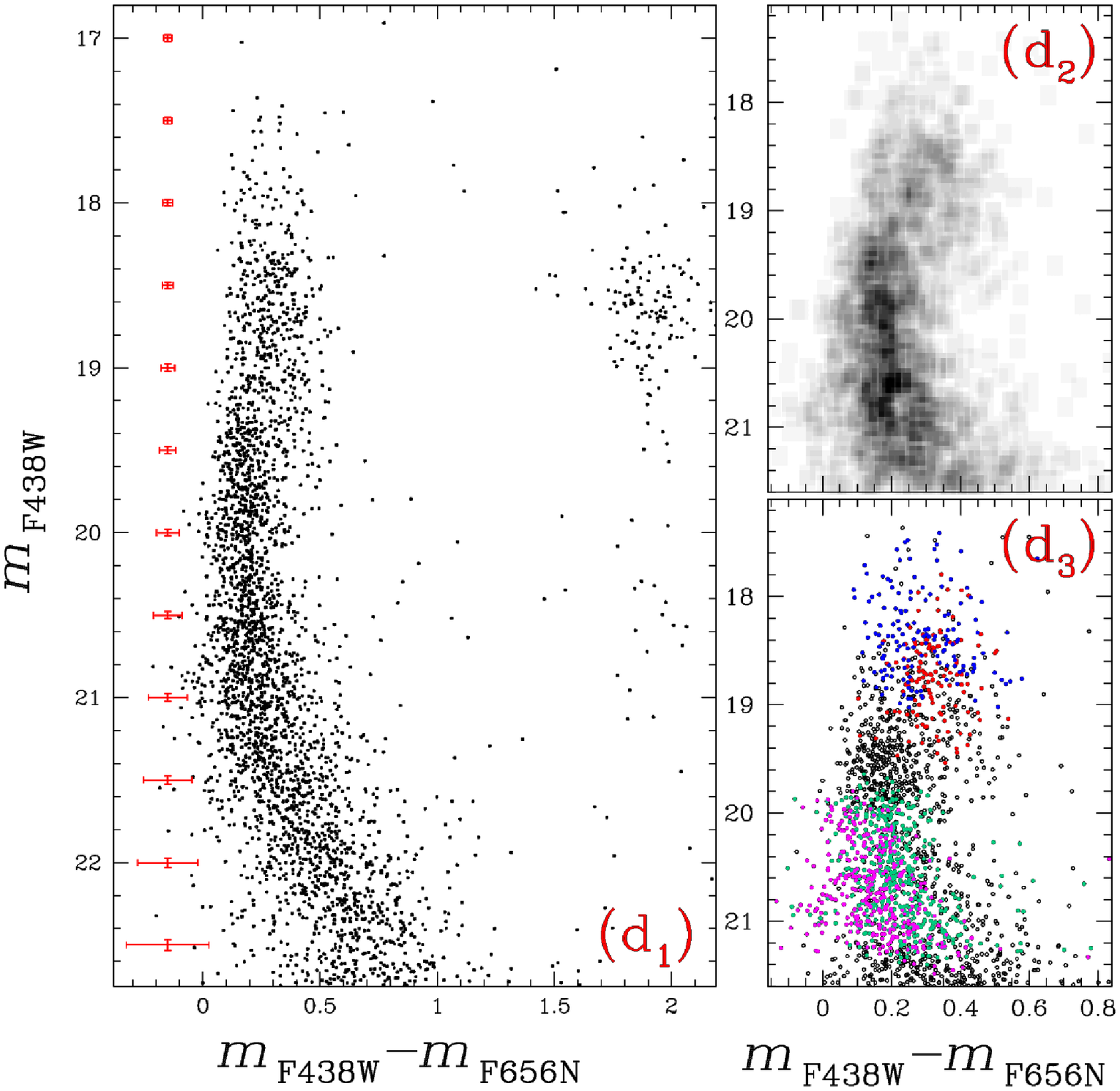} 
%/home/milone/MC/NGC1856/all/MATCH/figure/fig1.macro go2a go3  
%/home/milone/MC/NGC1856/all/MATCH/figure/fig1.macro go2b go2d 
 \caption{
Panel (a$_1$): $m_{\rm F336W}$ vs.\,$m_{\rm F336W}-m_{\rm F555W}$ CMD corrected for differential reddening for stars in the cluster field. Panel (a$_{2}$) shows the corresponding Hess diagram for stars around the MSTO. In panel (a$_{3}$) we have identified a group of bMS and a group of rMS stars that we have colored magenta and green, respectively, and a sample of bMSTO and a  of fMSTO colored blue and red, respectively. The same color codes have been used to represent the same stars in panels (b$_{3}$), (c$_{3}$), and (d$_{3}$). Panels (b$_{1,2,3}$), (c$_{1,2,3}$), and (d$_{1,2,3}$) are the same as panels (a$_{1,2,3}$) but for the reddening-free $m_{\rm F438W}+m_{\rm F814W}-1.838 m_{\rm F336W}$ vs.\,($m_{\rm F336W}-m_{\rm F438W}$)-1.0065 ($m_{\rm F438W}-m_{\rm F555W}$) pseudo CMD, and for the $m_{\rm F438W}$ vs.\,$m_{\rm F438W}-m_{\rm F814W}$ and $m_{\rm F438W}$ vs.\,$m_{\rm F438W}-m_{\rm F656N}$ CMDs corrected for differential reddening. } 
 \label{fig:pop2} 
\end{figure*} 
\end{centering} 
%%%%%%%%%%%%%%%%%%%%%%%%%%%%%%%%%%%%%%%%%%%%%%%%%%%%%%%%%%%%%%%%%%%%%%%%%%%%%%% 

In order to investigate the MSTO morphology of NGC\,1856 we have used the $m_{\rm F336W}$ vs.\,$m_{\rm F336W}-m_{\rm F555W}$ CMD for stars in the cluster field of view corrected for differential reddening (panel (a$_{1}$) of Fig.~\ref{fig:pop2}).  
A visual inspection of this CMD confirms that NGC\,1856 harbours an eMSTO. Furthermore, the MS reveals two main components below the TO, which are separated by $\sim$0.1 mag at $m_{\rm F336W} \sim$20.0 and merge together $\sim$1.5 magnitudes below. These features are highlighted by the Hess diagram in panel (a$_{2}$). 
     
In the CMD of Fig.~\ref{fig:pop2}a$_{3}$ we have arbitrarily selected two samples of bright MSTO (bMSTO) and faint MSTO (fMSTO) stars that we have colored in blue and red, respectively. Moreover we have used magenta and green color codes to represent two groups of blue MS (bMS) and red MS (rMS)  stars. 
 
To further demonstrate that the broadened MSTO and the double MS are not due to  differential reddening we have introduced the $m_{\rm F438W}+m_{\rm F814W}-1.838 m_{\rm F336W}$ vs.\,($m_{\rm F336W}-m_{\rm F438W}$)-1.0065 ($m_{\rm F438W}-m_{\rm F555W}$) shown in the panel (b$_1$) of Fig.~\ref{fig:pop2}.  
 The pseudo-color and pseudo magnitude used in this diagram are constructed in such a way that their total extinction equals to zero, thus making the diagram of Fig.~\ref{fig:pop2}b$_1$  reddening free. Indeed, by using the absorption coefficients provided by Aaron Dotter and listed above we derive: $A_{\rm F438W}+A_{\rm F814W}-1.838 A_{\rm F336W}$=($A_{\rm F336W}-A_{\rm F438W}$)-1.0065 ($A_{\rm F438W}-A_{\rm F555W}$)=0. 
 The fact that the MSTO and the upper MS of NGC\,1856 are more broadened than what is expected from observational errors further demonstrates that the broadening can not be due to residual differential reddening. 
  As noticed by the referee, the Hess diagram and the pseudo-CMD of panels b$_2$ and b$_3$ may reveal more than two sequences. Below $m_{\rm F438W}+m_{\rm F814W}-1.838 m_{\rm F336W}$$\sim$16.2, the rMS and the bMS are clearly visible, while at brighter luminosities, around the MSTO, there is some hint of a triple sequence.
Unfortunately the connection between the MSs and the MSTOs is not clear from the diagrams. A visual inspection of the Hess diagram of panel b$_2$ suggests that the rMS could further split into distinct MSs, which evolve into the bMSTO and the fMSTO. In this case the bMS should evolve into the bluest region of the bMSTO.
 As an alternative the bMS could cross the rMS and evolve into the fMSTO.
 In the next Sect.~\ref{subsec:age} and~\ref{subsec:CNO} we will further investigate these hypothesis.

\subsection{Spatial variations of photometric zero point} 
Apart from observational errors, spatial variations of the photometric zero point either due to differential reddening or to small uncertainties in the PSF model can broaden the sequences in the CMD and mimic multiple stellar populations (e.g.\,Anderson et al.\,2008). 
 To investigate this phenomenon, we exploit the $m_{\rm F438W}$ vs.\,$m_{\rm F438W}-m_{\rm F814W}$ and  $m_{\rm F438W}$ vs.\,$m_{\rm F438W}-m_{\rm F656N}$ CMDs and Hess diagrams plotted in the lower panels of Fig.~\ref{fig:pop2}. The comparison of the observed CMD of both panel (c$_{1}$) and (d$_{1}$) with the error bars plotted on the left reveals that the upper MS of NGC\,1856 is more broadened than expected from photometric errors only. 
 
To demonstrate that this broadening is not entirely due to photometric zero-point spatial variations, in the panels (c$_{3}$) and (d$_{3}$) of Fig.~\ref{fig:pop2} we have used the same colors introduced in Fig.~\ref{fig:pop2}a$_{3}$ to represent the bMSTO, fMSTO, bMS, and rMS stars defined therein. 
 As suggested by Anderson et al.\,(2009), if the color and magnitude broadening of the MS is entirely due to photometric errors, a star that is redder than the bulk of MSTO stars in $m_{\rm F336W}-m_{\rm F555W}$ has the same probability of being either redder or bluer in $m_{\rm F438W}-m_{\rm F814W}$ or $m_{\rm F438W}-m_{\rm F656N}$. Instead we notice that the two groups of bMSTO and fMSTO stars selected in the CMD of panel (a$_3$) are also well separated in the CMDs of panels (c$_{3}$) and (d$_{3}$).  This further demonstrates that the broadening of the MSTO can not be entirely due to measurement errors. 
 Similarly, from the $m_{\rm F336W}-m_{\rm F555W}$, the $m_{\rm F438W}-m_{\rm F814W}$, and the $m_{\rm F438W}-m_{\rm F656N}$ colors of the bMS and rMS stars we conclude that NGC\,1856 hosts at least two MS components fainter than the TO.  
 
%%%%%%%%%%%%%%%%%%%%%%%%%%%%%%%%%%%%%%%%%%%%%%%%%%%%%%%%%%%%%%%%%%%%%%%%%%%%%%% 
\subsection{Field stars}\label{sub:field} 
 
%%%%%%%%%%%%%%%%%%%%%%%%%%%%%%%%%%%%%% FIG 4 %%%%%%%%%%%%%%%%%%%%%%%%%%%%%%%%%%% 
\begin{centering} 
\begin{figure} 
 \includegraphics[width=9cm]{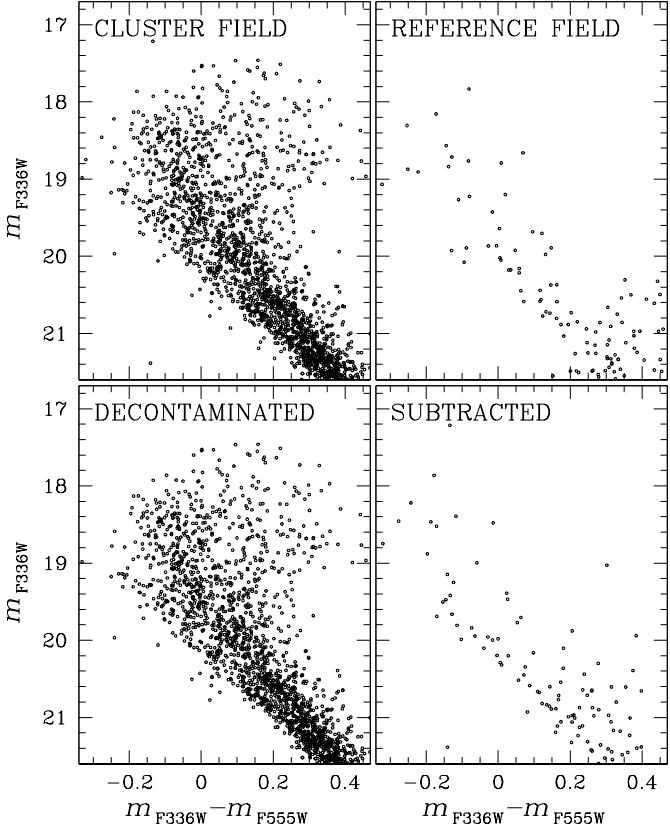} 
%/home/milone/MC/NGC1856/all/MATCH/figure/fig1.macro go4 
 \caption{\textit{Upper panels:} $m_{\rm F336W}$ vs.\,$m_{\rm F336W}-m_{\rm F555W}$ CMD of stars in the cluster field (left) and for stars in the reference field (right). \textit{Lower panels:} CMD for cluster-field stars after field stars have been statistically subtracted (left) and CMD of the subtracted stars (right).} 
 \label{fig:decon} 
\end{figure} 
\end{centering} 
%%%%%%%%%%%%%%%%%%%%%%%%%%%%%%%%%%%%%%%%%%%%%%%%%%%%%%%%%%%%%%%%%%%%%%%%%%%%%%% 
Contamination from background and foreground stars can be responsible for multiple sequences in the CMD. In this subsection we present the decontaminated cluster CMD to show that the observed eMSTO and the split MS can not be explained by field stars.
The best way to derive a CMD solely made of NGC\,1856 stars is by separating field stars from cluster members using stellar motions. 
 Unfortunately, the short baseline of our observations coupled with the large distance of the LMC prevents us to derive  proper motions that are accurate enough to reach this goal. 
As an alternative, we note that, due to the small size of the analyzed WFC3/UVIS field of view, we can assume that the distribution of field stars is homogeneous.   
 Therefore, to investigate whether the broadened MS of NGC\,1856 is due to field-star contamination or not, we follow the recipe of Paper\,I and II and compare the CMD of stars in the cluster field and in a reference field which cover the same area as the cluster field but includes only field stars.  
 This procedure is illustrated in Fig.~\ref{fig:decon}, where we show in the upper panels a zoom of the $m_{\rm F336W}$ vs.\,$m_{\rm F336W}-m_{\rm F555W}$ CMD around the MSTO of NGC\,1856 for stars in the cluster field (upper-left panel) and in the reference field (upper-right panel).  
  
We find that, in the interval of color and magnitude shown in Fig.~\ref{fig:decon} with  $-0.34 < m_{\rm F336W}-m_{\rm F555W} < 0.47$ and $16.7 < m_{\rm F336W} < 21.6$ , the cluster and the reference fields contain 2031 and 107 stars, respectively. 
This fact suggests that the field contamination in the CMD of NGC1856 is of the order of only 5\%.
 
 Then we have statistically subtracted the stars in the reference field from the CMD of stars in the cluster field. To do this we have defined for each star (i) in the reference field a distance\\ {\scriptsize $d_{\rm i}$=$\sqrt { (k ((m_{\rm F336W, cf}-m_{\rm F555W, cf})-(m_{\rm F336W, rf}^{\rm i}-m_{\rm F555W, rf}^{\rm i})))^{2}   +  (m_{\rm F336W, cf}-m_{\rm F336W, cf}^{\rm i})^{2}}$},\\ where $m_{\rm F336W, cf}$ and $m_{\rm F555W, cf}$ are the magnitudes of stars in the cluster field, while the magnitudes of stars in the reference field are indicated as $m_{\rm F336W, rf}$ and $m_{\rm F555W, rf}$. 
To account for the fact that the color of a star is better constrained than its magnitudes (see discussion in Gallart et al.\,2003) we have multiplied the color by a factor k=3.2, whose value has been determined as in Marino et al.\,(2014, see their Sect.~3.1).  
 Finally, we flagged the closest cluster-field star as a candidate to be subtracted.

The decontaminated CMD of NGC\,1856 plotted in the lower-left panel of Fig.~\ref{fig:decon} demonstrates that the eMSTO and the broad MS of NGC\,1856 are not due to field stars. For completeness, we plot the CMD of subtracted stars in the lower-right panel of Fig.~\ref{fig:decon}. 
 
\subsection{Non-interacting binaries}\label{binarie} 
 Paper\,I and II have shown that the CMD of all the analyzed young and intermediate-age star cluster in the LMC host a large fraction of non-interacting binaries, which ranges from $\sim$0.19, in the case of NGC\,1652 to $\sim$0.46 for NGC\,2108. 
 A visual inspection of the CMDs of Fig.~\ref{fig:reddening} and Fig.~\ref{fig:pop2}c$_1$ reveals an excess of stars on the red side of the MS and suggests that NGC\,1856 has a sizeable binary population. Since we expect that binaries can affect the distribution of colors and magnitudes of MS stars, we have investigated the effect of binaries on the morphology of the turn off.    
 To do this we first have determined the fraction of binaries along the MS (MS-MS binaries). Then we have simulated a CMD of NGC\,1856 which includes the derived binary fraction and have compared it with our observations.  

\begin{centering} 
\begin{figure*} 
 \includegraphics[width=12.5cm]{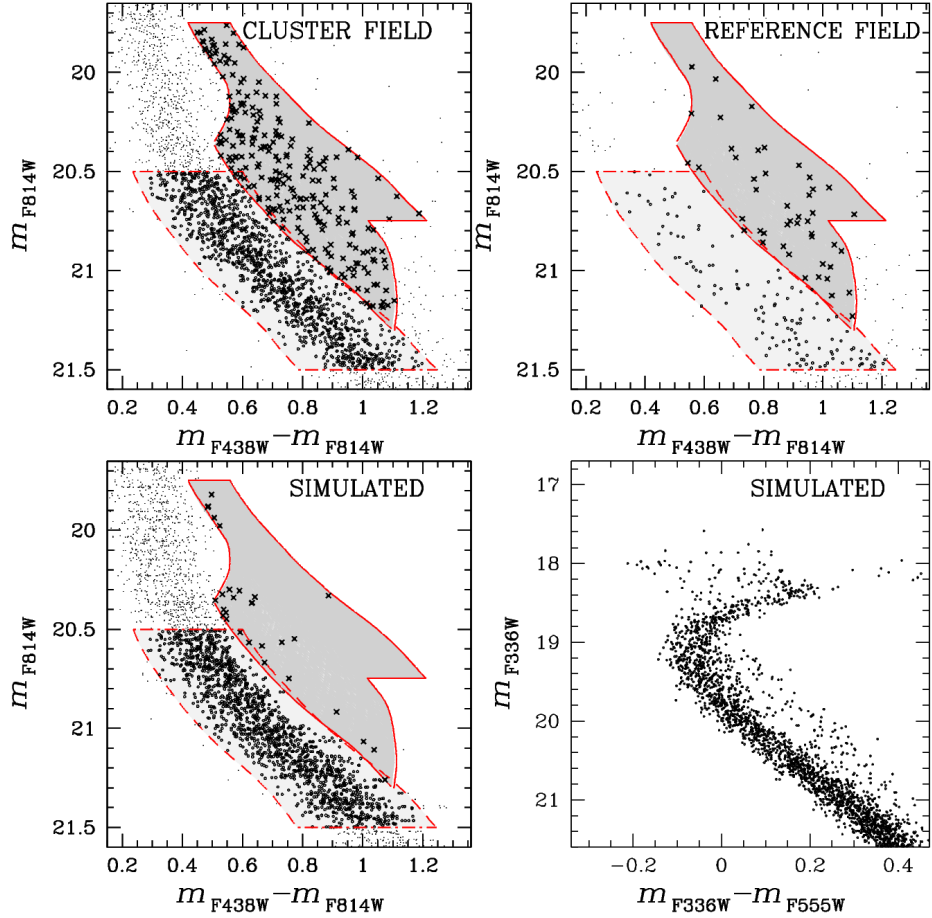} 
%/home/milone/MC/NGC1856/all/XKS/KSBI/macro/binarie/luoghi.macro fig 
 \caption{ $m_{\rm F814W}$ vs.\,$m_{\rm F438W}-m_{\rm F814W}$ CMD for stars in the cluster field (upper-left), in the reference field (upper-right), and simulated CMD (lower-left). Light- and dark-gray areas
mark the region A of the CMD used to estimate the binary fraction. Region B is the sub-region of A colored dark-gray. 
 Candidate single stars and binaries with q$<$0.5 in the analyzed interval of magnitude are represented with black points, while crosses represent candidate binaries with q$>$0.5. The simulated  $m_{\rm F336W}$ vs.\,$m_{\rm F336W}-m_{\rm F555W}$ CMD is plotted in the bottom-right panel. See text for details.} 
 \label{fig:binaries} 
\end{figure*} 
\end{centering}

To infer the fraction of MS-MS binaries in NGC\,1856 we have used the $m_{\rm F814W}$ vs.\,$m_{\rm F438W}-m_{\rm F814W}$ CMD,  where the binaries with $q>$0.5 are well separated from single MS stars, and followed the recipe of Milone et al.\,(2012). The quantity $q$ is the mass ratio of the two components of the binary.
Briefly, we have divided the CMD into two regions A, and B. Region A includes all the single stars and binary systems with a primary star with 20.5$<m_{\rm F814W}<$21.5, and includes both the light- and dark-shaded area of Fig.~\ref{fig:binaries}. Region B, is the dark-shaded portion of region A and includes binaries with $q>0.5$. 
 The fraction of binaries with a mass ratio $q>$0.5 has been calculated as 
\begin{center} 
$ 
f_{\rm BIN}^{\rm q>0.5}=\frac{N_{\rm CLUSTER}^{\rm B} - N_{\rm REFERENCE}^{\rm B}} 
                        {N_{\rm CLUSTER}^{\rm A} - N_{\rm REFERENCE}^{\rm A} } 
                 - \frac{N_{\rm ARTS}^{\rm B}} {N_{\rm ARTS}^{\rm A}} 
$ 
\end{center} 
where $N_{\rm CLUSTER}^{\rm A, B}$ are the number of stars, corrected for completeness, in the regions A and B of the CMD of stars in the cluster field plotted in the upper-left panel of Fig.~\ref{fig:binaries}. 
 $N_{\rm REFERENCE}^{\rm A, B}$ are the corresponding numbers for stars in the CMD of stars in the reference field (upper-right panel of Fig.~\ref{fig:binaries}), while $N_{\rm ARTS}^{\rm A, B}$ are the numbers of stars in the regions A and B of the CMD made of artificial stars (lower-left panel of Fig.~\ref{fig:binaries}, see Milone et al.\,2012 for details). 
 
 We conclude that NGC\,1856 hosts a large fraction of binaries, in close analogy with what is observed in other MC clusters. We find $f_{\rm BIN}^{\rm q>0.5}=$0.147$\pm$0.015 in the cluster field and infer a total fraction of binaries $f_{\rm BIN}^{\rm TOT}$=0.294$\pm$0.031 by assuming a flat mass-ratio distribution for binaries with $q<$0.5.  
 In the lower-right panel of Fig.~\ref{fig:binaries} we have used artificial stars and BaSTI isochrones\footnote{http://www.oa-teramo.inaf.it/BASTI} (Pietrinferni et al.\,2004, 2006, 2009) to simulate the $m_{\rm F336W}$ vs.\,$m_{\rm F336W}-m_{\rm F555W}$ CMD of a simple stellar population with age of 
320 Myr and Z=0.008 (see Sect.~\ref{subsec:age} for an age estimate) and by assuming the same fraction of binaries above derived. 
 From the comparison of the simulated CMD and the observed one plotted in Fig.~\ref{fig:pop2}a$_{1}$   we conclude that non-interacting binaries alone are not able to reproduce neither the eMSTO nor the double MS observed in NGC\,1856. 
 
  The evidence of an eMSTO in NGC\,1856 is in disagreement with the conclusions by Bastian \& Silva Villa (2013) who have analyzed WFPC2 photometry in F555W and F814W of NGC\,1856 and found no evidence for an eMSTO in this cluster.
 Their work is based on the catalog by Brocato et al.\,(2001) who provided high quality photometry for the standard available at that time. 
 Our finding of an eMSTO and a split MS in NGC\,1856 is likely due, in part, to the high quality of the photometry obtained from images taken with the new WFC3/UVIS and ACS/WFC cameras of {\it HST} and by using the recent techniques developed by Jay Anderson and collaborators (see Sect.~\ref{sec:data} for details.)

Moreover, we have shown that the eMSTO is better visible when we use colors involving the F336W filter (like F336W-F555W and F336W-F814W colors colors, see Fig.~\ref{fig:reddening} and~\ref{fig:pop2}). Therefore the F336W photometry used in this paper is more efficient than F555W and F814W to detect the eMSTO in young clusters.

\section{The double MS of NGC\,1856} 
\begin{centering} 
\begin{figure*} 
 \includegraphics[width=12.5cm]{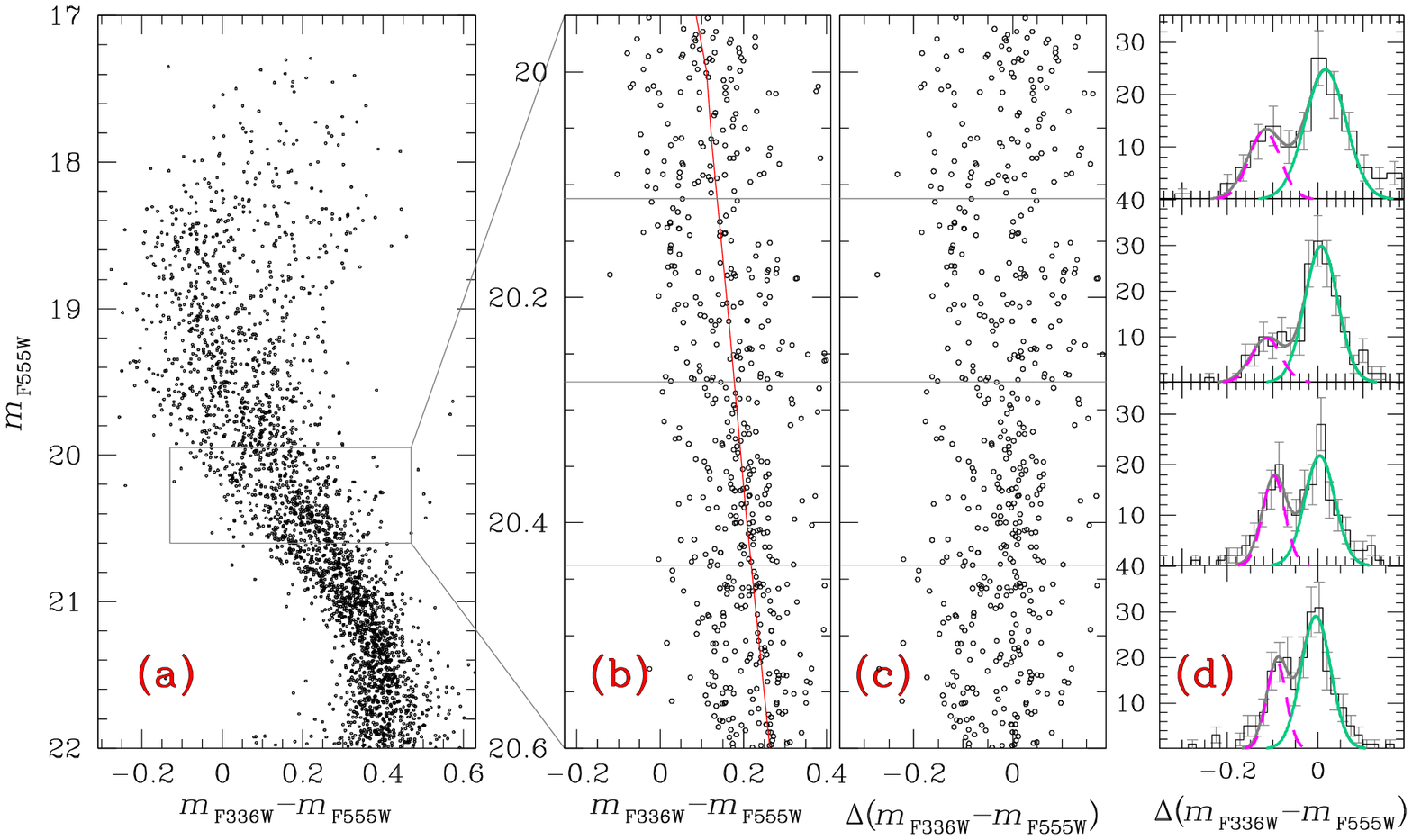} 
%/home/milone/MC/NGC1856/all/MATCH/figure/fig.macro goms 
 \caption{$m_{\rm F555W}$ vs.\,$m_{\rm F336W}-m_{\rm F555W}$ CMD of stars in the cluster field (panel a). Panel b shows a zoom of the region where the MS split is more clearly visible. The red line overimposed on the CMD is the fiducial line of the rMS. The verticalized $m_{\rm F555W}$ vs.\,$\Delta$($m_{\rm F336W}-m_{\rm F555W}$) diagram is plotted in panel c, while panels d show the histogram of the $\Delta$($m_{\rm F336W}-m_{\rm F555W}$) 
distribution for stars in the four magnitude intervals delimited by gray-horizontal lines in panels b and c. The least-squares best-fit bi-Gaussian function is colored gray, while the two components are represented with magenta and aqua color codes.} 
 \label{fig:dms} 
\end{figure*} 
\end{centering}

Section~\ref{sec:cmd} has provided evidence that, below the MSTO, the MS of NGC\,1856 is split, and that such splitting is real. 
 To our knowledge, after the case of NGC\,1844 (see Paper\,II), this is the second time that a double or broad MS in a young star cluster has been detected. This finding deserves further analysis to determine the fraction of rMS and bMS stars. In order to do this, we have applied to NGC\,1856 a procedure which has been widely used in previous studies of multiple stellar populations in GCs (e.g.\,Piotto et al.\,2007) and is illustrated in Fig.~\ref{fig:dms}.   
 
The double MS of NGC\,1856 is clearly visible in the panel a of Fig.~\ref{fig:dms}  where we show the $m_{\rm F555W}$ vs.\,$m_{\rm F336W}-m_{\rm F555W}$ CMD of stars in the cluster field. Only stars within the gray rectangle over the magnitude interval 19.9$<m_{\rm F555W}<$20.6, where the MS split is most prominent, are used in the following analysis. This region of the CMD is zoomed in panel b where we also plot the fiducial line of the red MS. 
 To determine the line we selected a sample of red MS stars  by eye  and calculated the median color and magnitudes of these stars in intervals of 0.2 mag in the F555W band. These median points have been interpolated with a cubic spline.  
The verticalized  $m_{\rm F555W}$ vs.\,$\Delta$($m_{\rm F336W}-m_{\rm F555W}$) CMD shown in panel c of Fig.~\ref{fig:dms} is obtained by subtracting the color of the fiducial at the corresponding $m_{\rm F555W}$ from the color of each star. 
 The color separation between bMS and rMS increases with luminosity and ranges from $\sim -$0.1 at $m_{\rm F555W}$=20.6 to $\sim -$0.15 at $m_{\rm F555W}$=19.9.  
 The  distribution  of $\Delta$($m_{\rm F336W}-m_{\rm F555W}$) is plotted
in panels d in four bins of $m_{\rm F555W}$. The histogram distributions exhibit two peaks that we have fitted with two Gaussians (magenta for the bMS and aqua for the rMS). From the area under the four Gaussian profiles we infer that the bMS and the rMS host 33$\pm$5\% and 67$\pm$5\% of MS stars, respectively, in the selected magnitude interval.  
 
%%%%%%%%%%%%%%%%%%%%%%%%%%%%%%%%%%%%%% FIG 3 %%%%%%%%%%%%%%%%%%%%%%%%%%%%%%%%%%% 
%%%%%%%%%%%%%%%%%%%%%%%%%%%%%%%%%%%%%%%%%%%%%%%%%%%%%%%%%%%%%%%%%%%%%%%%%%%%%%% 
 
\section{Possible origin of the MS split and of the eMSTO}\label{sec:teoria} 
In order to shed light on the physical reasons at the basis of the double MS and the eMSTO of NGC 1856, we have compared the observed CMD
with isochrones from the BaSTI archive. To do this, we have used isochrones which account for the core-convective overshoot during the central H-burning stage as in Pietrinferni et al.\,(2004). 
 Specifically, in Sect.~\ref{subsec:age} we test the hypothesis that the 
spread around the MSTO and in the MS are due to age variation only, while, in Sect.~\ref{subsec:CNO}, we investigate possible effects of 
C$+$N$+$O and iron variation on the CMD morphology. 

\subsection{Age and age spread}\label{subsec:age} 
Several authors suggested that the eMSTO of intermediate-age clusters in the MCs is due to a prolonged star-formation history (e.g.\,Mackey et al.\,2008; Paper\,I; Goudfrooij et al.\,2011). In this section we investigate the hypothesis that the eMSTO of the young cluster NGC\,1856 is due to multiple stellar generations with the same chemical composition and infer the consequent spread in age. 
   
\begin{centering} 
\begin{figure*} 
 \includegraphics[width=12.5cm]{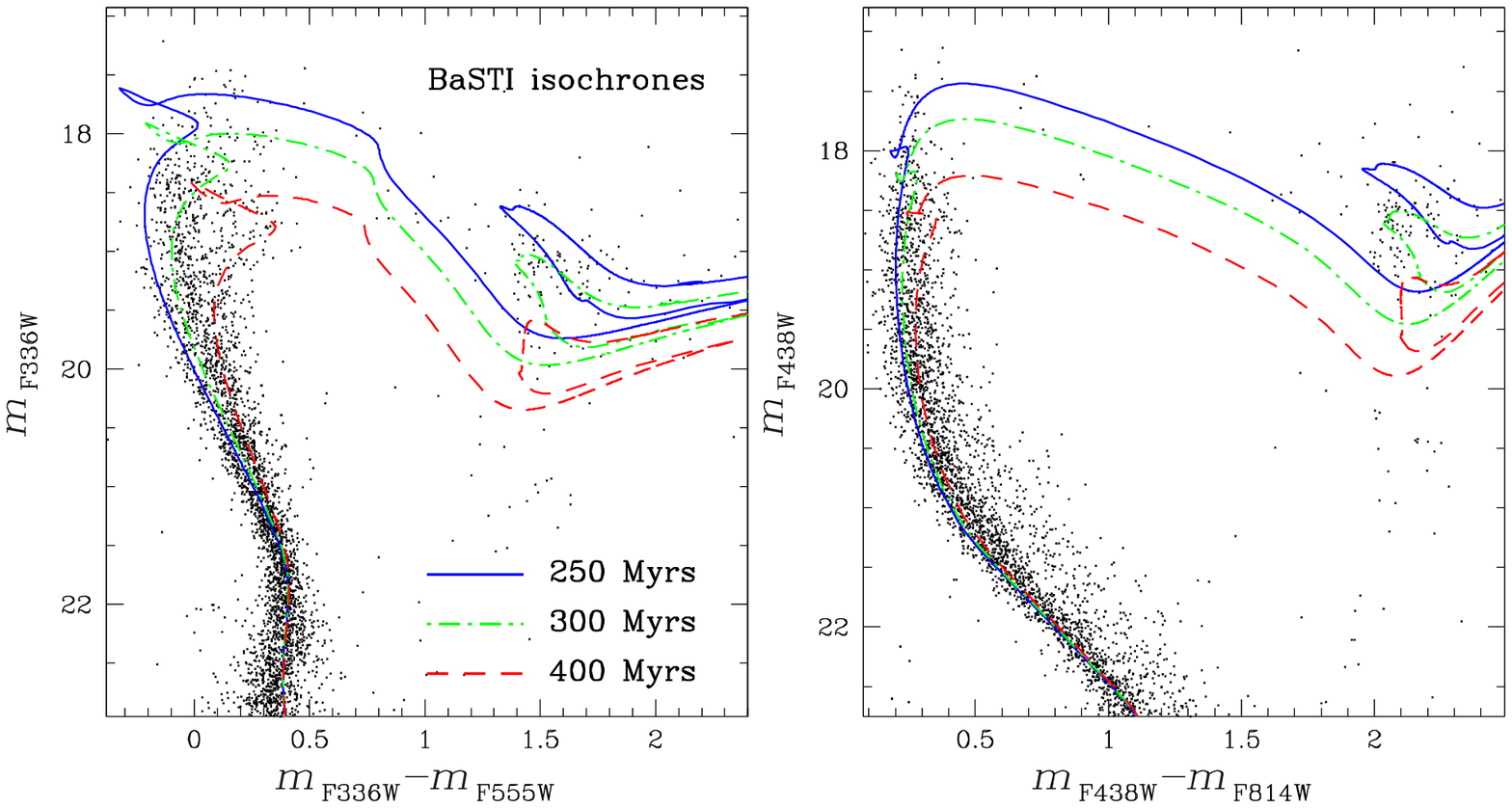} 
%/home/milone/MC/NGC1856/all/MATCH/figure/fig.macro go5b 
 \caption{The three isochrones from the BaSTI database which
provide the best match with the observed CMDs. Specifically, the blue and the red isochrones match
the bluer and redder envelope of the eMSTO of NGC\,1856, respectively, while the green ones fit the average distribution 
of cluster stars in the CMD. 
See text for details.} 
 \label{fig:age} 
\end{figure*} 
\end{centering}

Figure~\ref{fig:age} shows that, adopting a distance modulus ($m-M$)$_{\rm 0}$=18.35, a reddening $E$($B-V$)=0.15, a metallicity Z=0.008, and [$\alpha$/Fe]=0
the average distribution of stars in the $m_{\rm F336W}$ vs.\,$m_{\rm F336W}-m_{\rm F555W}$ and in the  $m_{\rm F438W}$ vs.\,$m_{\rm F438W}-m_{\rm F814W}$ CMDs can be reproduced by assuming a 300~Myr age, while the blue and the red envelopes of the MS and the MSTO are well matched by a 250-Myr and 400-Myr old isochrone, respectively.

%%%%%%%%%%%%%%%%%%%%%%%%%%%%%%%%%%%%%% FIG 3 %%%%%%%%%%%%%%%%%%%%%%%%%%%%%%%%%%% 
%%%%%%%%%%%%%%%%%%%%%%%%%%%%%%%%%%%%%%%%%%%%%%%%%%%%%%%%%%%%%%%%%%%%%%%%%%%%%%% 
 
\begin{centering} 
\begin{figure*} 
 \includegraphics[width=12.5cm]{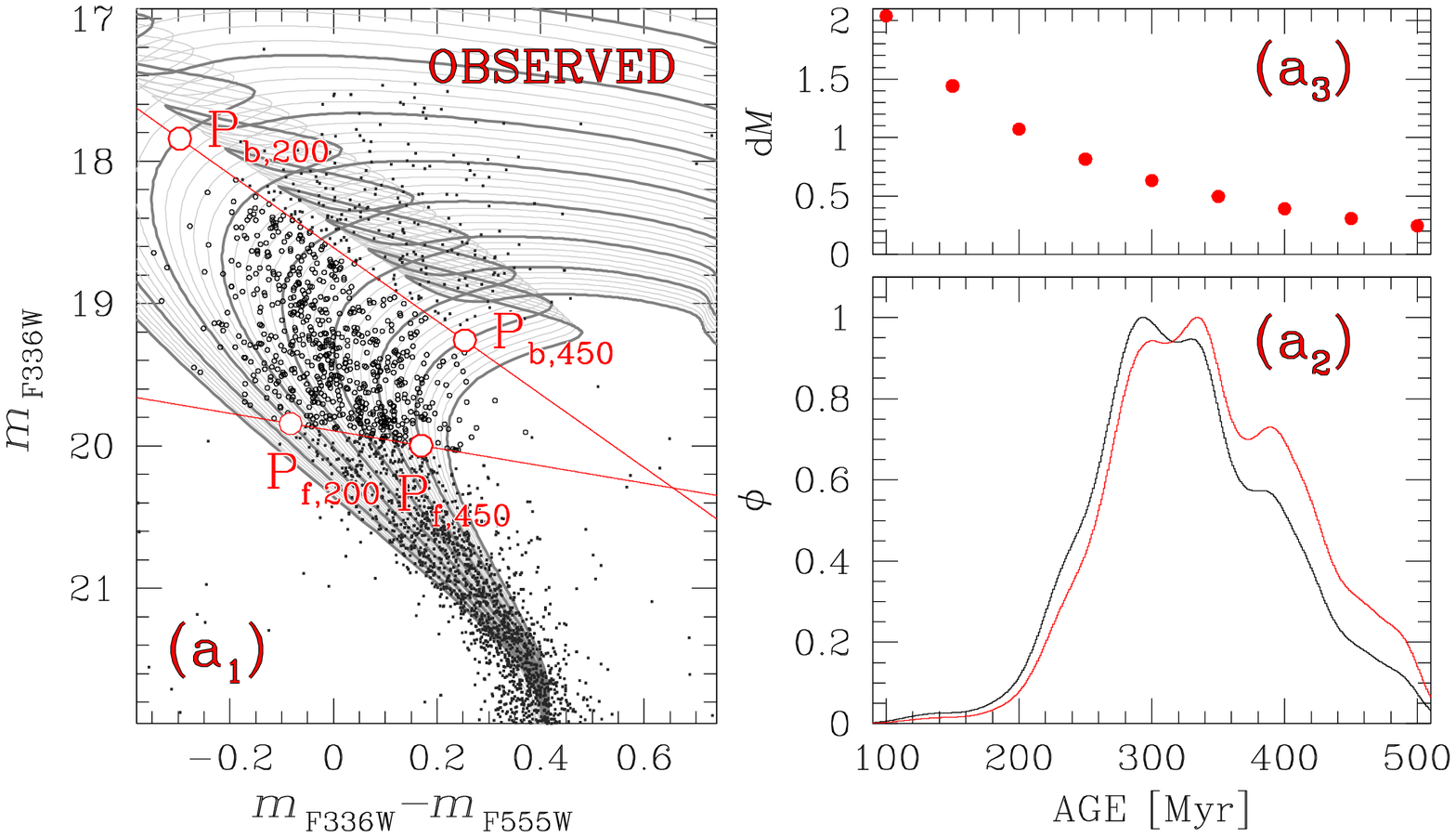} 
%/home/milone/MC/NGC1856/all/MATCH/figure/fig.macro fig (go6 go6b) 
 \includegraphics[width=12.5cm]{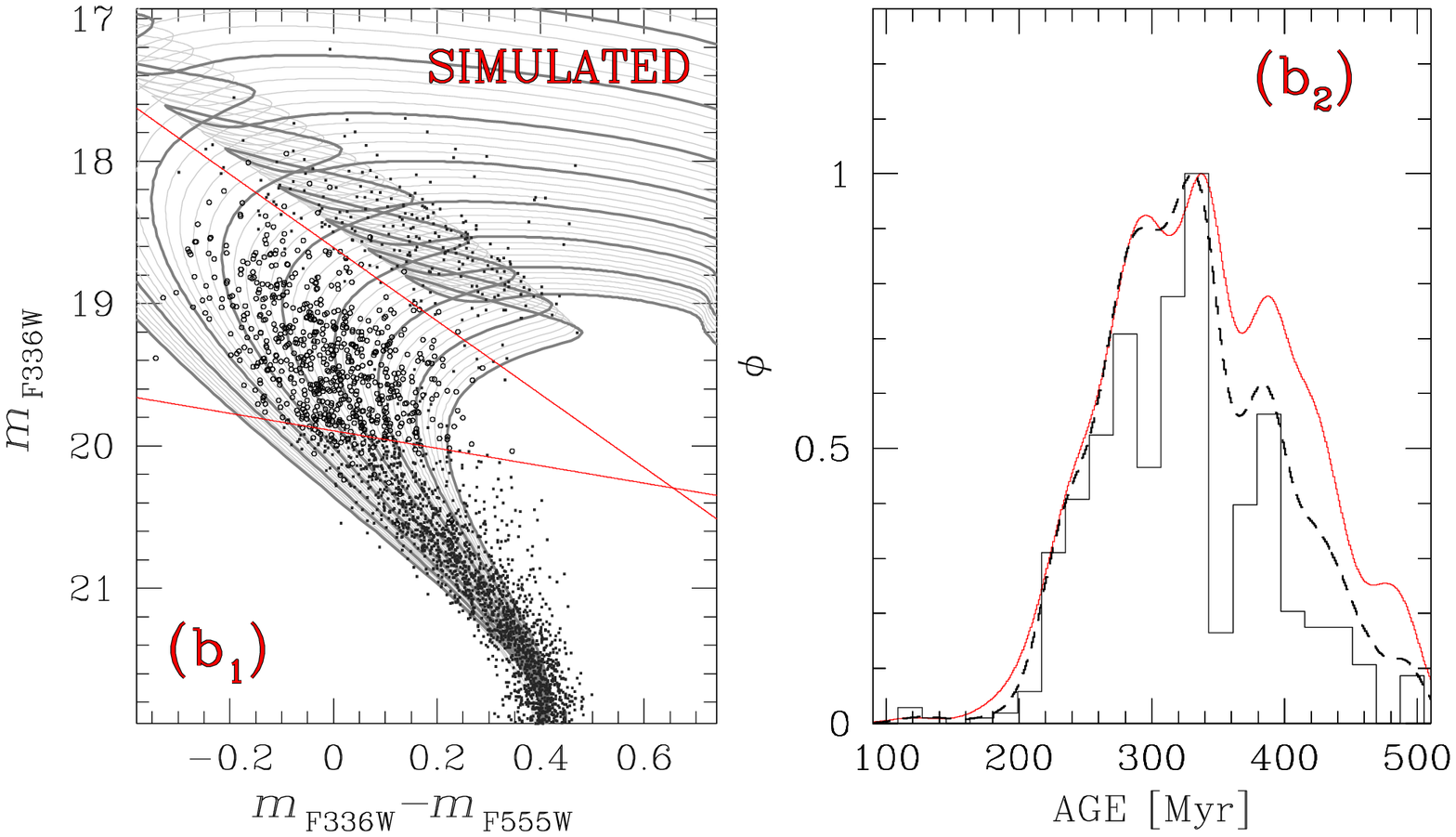} 
%/home/milone/MC/NGC1856/all/MATCH/figure/fig.macro fig2 (go6a go6ab) 
 \caption{ This figure illustrates the method to derive the age spread of NGC\,1856. Panel (a$_{1}$) reproduces the $m_{\rm F336W}$ vs.\,$m_{\rm F336W}-m_{\rm F555W}$ CMD of Fig.~\ref{fig:reddening} with overimposed a grid of isochrones from the BaSTI database. Isochrones with ages from 200 to 500 Myrs in steps of 50 Myrs are represented with thick gray lines, while the isochrones plotted with thin lines are separated by 10 Myrs.  The two red lines delimit the CMD region where the color and magnitude broadening is more evident. Only stars in this region, which are plotted with small black circles, are used to derive the distribution of ages. We have indicated with large red circles the intersections ($P_{\rm b,f,200}$, and $P_{\rm b,f,450}$) between the 200- and 450-Myr old isochrones and the two lines.  
  The kernel-density distribution of the ages plotted in panel (a$_{2}$) and derived for stars between the two red lines is represented with a black line, while the red lines represent the global distribution inferred for NGC\,1856. 
In panel (a$_{3}$), we have plotted the mass difference d$\mathcal{M}$ as a function of the age (see text for details).  
Panel (b$_1$) shows the simulated $m_{\rm F336W}$ vs.\,$m_{\rm F336W}-m_{\rm F555W}$ CMD obtained by assuming the age distribution corresponding to the black histogram of panel (b$_2)$. The black-dashed line indicates the kernel-density distribution of the ages of simulated stars, while the red-continuous line is the distribution inferred from the analysis of the simulated CMD of panel (b$_{1}$). 
} 
 \label{fig:agespread} 
\end{figure*} 
\end{centering}

The comparison between isochrones and observations provides a rough estimate of the duration of the star formation in NGC\,1856 (in the hypothesis of prolonged or multiple star formation). 
To determine a more accurate age distribution of stars in NGC\,1856 we have used the procedure illustrated in Fig.~\ref{fig:agespread}.  
We started by superimposing two red lines on the $m_{\rm F336W}$ vs.\,$m_{\rm F336W}-m_{\rm F555W}$ CMD, as shown in panel (a$_{1}$). These two lines have been drawn by hand with the criterion of selecting the region around the MSTO where the color and magnitude spread due to age variation is more clearly distinguishable. 
 Only the stars between the two lines and belonging
to the decontaminated CMD of Sect.~\ref{sub:field} (black circles in Fig.~\ref{fig:agespread}a$_{1}$) have been used to estimate
the age distribution. 
  
To do this, we have superimposed on the CMD the grid of isochrones available from the BaSTI database for the same values of metallicity, [$\alpha$/Fe], distance and reddening as above described. 
The available isochrones
from the BaSTI webpage -- represented in Fig.~\ref{fig:agespread} with thick gray lines -- have ages between 100 Myr and 550 Myr in steps of 50 Myrs. The isochrones represented with thin gray lines are separated by 10 Myr and have been obtained by linearly interpolating the BaSTI isochrones.   
 
In order to estimate the age of a given star, j, we started by calculating the distance in the CMD between star j and all isochrones. 
Then we have defined the quantities D$_{1}$=d$_{1}$/(d$_{1}$+d$_{2}$) and D$_{2}$=d$_{2}$/(d$_{1}$+d$_{2}$), where 
d$_{1}$ and d$_{2}$ are the distances from the two closest isochrones of age t$_{1}$ and t$_{2}$ of a given star.
The resulting
 best age estimate is: 
$t_{\rm j}= t_{1} D_{2} + t_{2} D_{1}$

To derive the kernel-density distribution of ages for the analyzed stars, we assumed a Gaussian kernel with dispersion equal to the observational error $\sigma$. Then we have divided the age interval 100$<$t$<$550 Myrs into bins of 1 Myr and, for each bin, we have calculated the quantity $\phi_{\rm i}=\sum_{\rm j=1,}^{\rm NTOT}{\frac{1}{c_{\rm j}} exp \frac{-(t_{\rm i}-t_{\rm j})^{2}}{2 \sigma^{2}}}$, where $c_{\rm j}$ is the completeness of the star j, and NTOT=1914 is the number of analyzed stars between the two red lines. The normalized age distribution is represented with a black line in panel (a$_{2}$) of Fig.~\ref{fig:agespread}.

 Each isochrone of age t intersects the two red lines of Fig.~\ref{fig:agespread}a at two points, $P_{\rm b,t}$ and $P_{\rm f,t}$, which correspond to mass $\mathcal{M}_{\rm b}$ and $\mathcal{M}_{\rm f}$.  
 As an example, in Fig.~\ref{fig:agespread}a we have indicated with large red circles the points corresponding to the 200-Myr and 450-Myr old isochrones. 
 The mass interval d$\mathcal{M}$=$\mathcal{M}_{\rm b}-{M}_{\rm f}$ changes from one isochrone to another as shown in the panel (a$_{3}$) of Fig.~\ref{fig:agespread}, where we plot d$\mathcal{M}$ as a function of the age of the isochrone. 
 
 As a consequence, the distribution $\phi_{\rm i}$ above obtained
clearly depends on our choice of the CMD region used to determine it.  
To remove such dependence from the derived age distribution, we have calculated the quantity $\phi_{\rm i}^{*}=\sum_{\rm j=1,}^{\rm NTOT}{\frac{1}{c_{\rm j} d\mathcal{M}_{\rm j}} exp \frac{-(t_{\rm i}-t_{\rm j})^{2}}{2 \sigma^{2}}}$, where $ d\mathcal{M}_{\rm j}$ is the mass interval corresponding to star j. The normalized age  distribution is represented with a red line in the panel (a$_{2}$) of Fig.~\ref{fig:agespread}. 
 
In order to estimate the observed age spread we have calculated the quantities age$_{10}$ and age$_{90}$ defined as the age values that satisfy the conditions: 
 $\sum_{\rm i=100,}^{\rm age_{10}} \phi_{\rm i}^{*} =\frac{10}{100} \sum_{\rm i=100,}^{550} \phi_{\rm i}^{*}$ and $\sum_{\rm i=100,}^{\rm age_{90}} \phi_{\rm i}^{*} =\frac{90}{100} \sum_{\rm i=100,}^{550} \phi_{\rm i}^{*}$, respectively, and defined the quantity $\sigma_{80, \rm OBS}$=age$_{90}$-age$_{10}$. Similarly, we have defined the quantity $\sigma_{68, \rm OBS}$. 
We obtained  $\sigma_{80, \rm OBS}$=185 Myr and $\sigma_{68, \rm OBS}$=146 Myr. 
 
Note that the age spread above estimated is not the intrinsic age dispersion. 
 In fact the measured age spread ($\sigma_{80(68), \rm OBS}$) results from the combination of the intrinsic age dispersion ($\sigma_{80(68), \rm{age}}$) and the broadening due to measurement errors and binary stars ($\sigma_{80(68), \rm ERR}$), i.e.\,{\scriptsize $\sigma_{80(68), \rm{age}}=\sqrt{\sigma_{80(68), \rm OBS}^{2}-\sigma_{80(68), \rm ERR}^{2}}$}.

To infer $\sigma_{80(68), \rm ERR}$ and determine the intrinsic age spread we have used AS tests
and simulated the $m_{\rm F336W}$ vs.\,$m_{\rm F336W}-m_{\rm F555W}$ CMD plotted in panel (b$_1$) of Fig.~\ref{fig:agespread}. 
We have followed the recipe described in detail in Paper\,I (see their Sect.~6.2). We used the same metallicity, $\alpha$ abundance, distance modulus and reddening adopted for the isochrone fit of Fig.~\ref{fig:age} 
and the same fraction of binaries as derived in Sect.~\ref{binarie}. We assumed the age distribution represented by the black histogram in panel  (b$_2$) of Fig.~\ref{fig:agespread} (the dashed black line represents the corresponding kernel-density distribution of the input ages for our simulation). 

We have applied to the simulated CMD the same procedure as used for the observed CMD to determine the age spread and derived the red line of panel (b$_2$).  
Note that the age distribution adopted as input to generate the simulated CMD [black histogram in panel (b$_2$)] has been derived with the criteria that the kernel-density age distribution derived from the simulated diagram [red line in panel (b$_2$)] reproduces the age distribution obtained from the observed CMD [(a$_2$), red line].
 We emphasize that the detailed morphology of the histogram representing the input age distribution for the simulated CMD could have been different. Here, we are interested in estimating the intrinsic age dispersion $\sigma_{68, \rm{age}}$ and $\sigma_{80, \rm{age}}$ not the detailed distribution of the  ages of the  stars in NGC\,1856. 
 What is relevant from our analysis is the similarity of the observed and simulated red lines in panels (a$_2$) and (b$_2$) of Fig.~\ref{fig:agespread}.

Also for the simulated CMD we defined a $\sigma_{80, \rm SIM}$ and $\sigma_{68, \rm SIM}$, in close analogy with what was done for the observed CMD. We obtain $\sigma_{80, \rm SIM}$=182 Myr and $\sigma_{68, \rm SIM}$=145 Myr. 
Similarly, the difference between the 90$^{\rm th}$ and the 10$^{\rm th}$  and the 84$^{\rm th}$ and the 16$^{\rm th}$ percentile of the age distribution used as input of our simulation [black histogram of panel (b$_2$)]  have been called $\sigma_{80, \rm INP}$ and $\sigma_{68, \rm INP}$, respectively. 
We have found $\sigma_{80, \rm INP}$=167 Myr and $\sigma_{68, \rm INP}$=138 Myr. Therefore the contribute of measurement errors and binary stars to the observed age spread is
{\scriptsize $\sigma_{80, \rm{ERR}}=\sqrt{\sigma_{80, \rm SIM}^{2}-\sigma_{80, \rm INP}^{2}}$}=72 Myr and {\scriptsize $\sigma_{68, \rm{ERR}}=\sqrt{\sigma_{68, \rm SIM}^{2}-\sigma_{68, \rm INP}^{2}}$}=45 Myr.
 
At this point we have estimated  the intrinsic age dispersion as 
 {\scriptsize $\sigma_{80, \rm{age}}=\sqrt{\sigma_{80, \rm OBS}^{2}-\sigma_{80, \rm ERR}^{2}}$}=174$\pm$31 Myr and {\scriptsize $\sigma_{68, \rm{age}}=\sqrt{\sigma_{68, \rm OBS}^{2}-\sigma_{68, \rm ERR}^{2}}$}=139$\pm$22 Myr.  
Uncertainties on $\sigma_{68}$ and $\sigma_{80}$ are estimated  by bootstrapping with replacements performed 1,000 times on both the observed and the simulated age distributions. 
 The errors indicate one standard deviation (68.27$^{\rm th}$ percentile) of the bootstrapped measurements.

%%%%%%%%%%%%%%%%%%%%%%%%%%%%%%%%%%%%%% FIG 6 %%%%%%%%%%%%%%%%%%%%%%%%%%%%%%%%%%% 
%%%%%%%%%%%%%%%%%%%%%%%%%%%%%%%%%%%%%%%%%%%%%%%%%%%%%%%%%%%%%%%%%%%%%%%%%%%%%%% 

\subsection{C$+$N$+$O and metallicity.}\label{subsec:CNO}
 Multiple stellar populations in nearly all Galactic GCs are characterized by star-to-star variation in several light elements, like C, N, Na, and O (e.g.\,Kraft\,1994; Gratton et al.\,2004, 2012; Carretta et al.\,2009). Moreover stars with different light-element abundance populate distinct sequences in CMDs built with ultraviolet magnitudes (e.g.\,Marino et al.\,2008; Yong et al.\,2008; Sbordone et al.\,2011; Dotter et al.\,2015). 
In some massive Galactic GCs, there are also internal variations in iron and in the overall C$+$N$+$O  abundance (e.g.\,Norris \& Da Costa\,1995; Cassisi et al.\,2008; Marino et al.\,2009; Ventura et al.,2009; Yong et al.\,2014) which are responsible for multiple sequences in the CMD (e.g.\,Piotto et al.\,2005; Marino et al.\,2011; Milone et al.\,2015). 
 
\begin{centering} 
\begin{figure*} 
 \includegraphics[width=12.5cm]{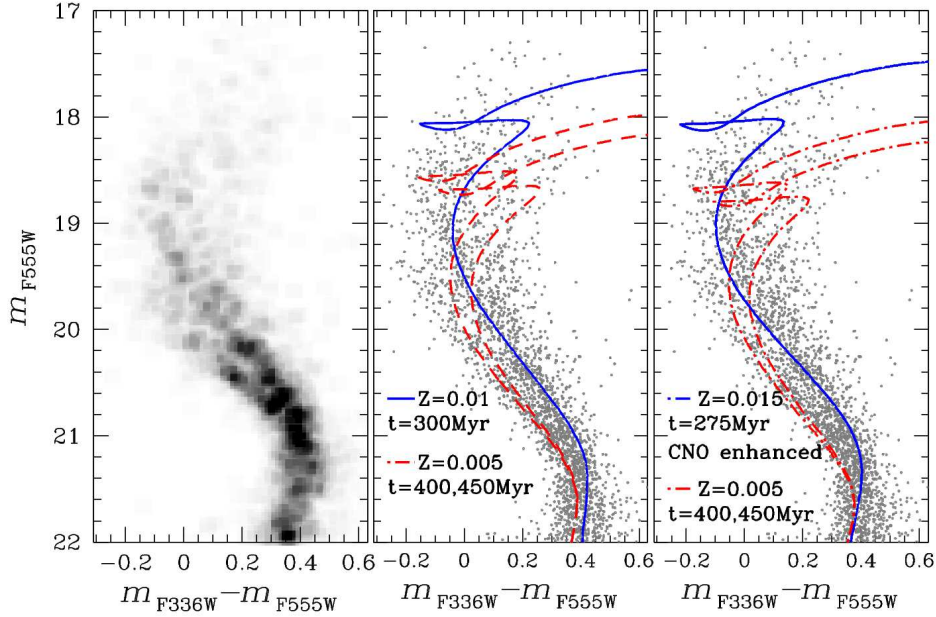} 
%/home/milone/MC/NGC1856/all/MATCH/figure/fig.macro gomsiso2 
 \caption{\textit{Left panel:} Hess diagram of the $m_{\rm F555W}$ vs.\,$m_{\rm F336W}-m_{\rm F555W}$ CMD of Fig.~\ref {fig:dms}. \textit{Middle panel:}  Solar scaled isochrones with different metallicity and age  from the BaSTI database are superimposed on the observed CMD.  \textit{Right panel:} As in the middle panel but for alpha-enhanced isochrones. In this case the isochrone with Z=0.015 is also CNO enhanced. See text for details.} 
 \label{fig:CNO} 
\end{figure*} 
\end{centering}

  In absence of a detailed study on the chemical-composition of NGC\,1856, in this subsection we investigate a possible effect of C$+$N$+$O and iron variation on the CMD morphology. 
 First of all we note that a visual inspection of the $m_{\rm F555W}$ vs.\,$m_{\rm F336W}-m_{\rm F555W}$ Hess diagram of stars in the cluster field plotted in the left panel of Fig.~\ref{fig:CNO} could suggest that the rMS and bMSTO are connected and that the bMS evolve into the fMSTO.

The middle panel of Fig.~\ref{fig:CNO} shows that a solar-scaled isochrone for Z=0.01, [Fe/H]=$-$0.25, and age, t=300\,Myr, matches the bulk of stars in NGC\,1856 and reproduces the rMS and the bMSTO. The same figure shows that the bMS and the fMSTO are consistent with  solar-scaled, metal poor isochrones with Z=0.005, [Fe/H]=$-$0.6 and age of 400-450\,Myr, although the fit is quite poor.

In the right panel of Fig.~\ref{fig:CNO} we have fitted the rMS and the bMSTO with an $\alpha$-enhanced isochrone with Z=0.015, [Fe/H]=$-$0.7, age of 275\,Myr and with a factor of $\sim$2 enhancement in CNO-element sum (blue line). The red isochrones correspond to $\alpha$-enhanced models for Z=0.005, [Fe/H]=$-$0.6 and age of 400-450\,Myr.
 
In summary, the comparison between the observed CMD and the isochrones of  Fig.~\ref{fig:CNO} shows that the bMS and the fMSTO are well fitted by an old population with [Fe/H]=$-$0.6, solar-scaled CNO, and age of  $\sim$400\,Myr, while the rMS and the bMSTO are consistent with a $\sim$300\,Myr-old stellar population. The later can be either enhanced in [Fe/H] by $\sim$0.4 dex or can have almost the same iron content as the old population but is enhanced in C$+$N$+$O by about a factor of two.
In all cases, even assuming variations in C, N, O and Fe we need an age spread of $\sim$150 Myr in order to reproduce the broadened distribution
of the upper MS and eMSTO.

 Finally, we enphasize that the analysis above only provides an attempt to interpret the eMSTO and the split MS of NGC\,1856 and that the present dataset does not allow us to to establish whether NGC\,1856 has homogeneous C$+$N$+$O and iron. Spectroscopy of NGC\,1856 stars is mandatory to properly contraint the chemical composition of this cluster.
%
%
%%%%%%%%%%%%%%%%%%%%%%%%%%%%%%%%%%%%%% FIG 3 %%%%%%%%%%%%%%%%%%%%%%%%%%%%%%%%%%% 
%%%%%%%%%%%%%%%%%%%%%%%%%%%%%%%%%%%%%%%%%%%%%%%%%%%%%%%%%%%%%%%%%%%%%%%%%%%%%%% 
 
\subsection{The extended red clump.}\label{subsec:bump}
\begin{centering} 
\begin{figure*} 
 \includegraphics[width=12.5cm]{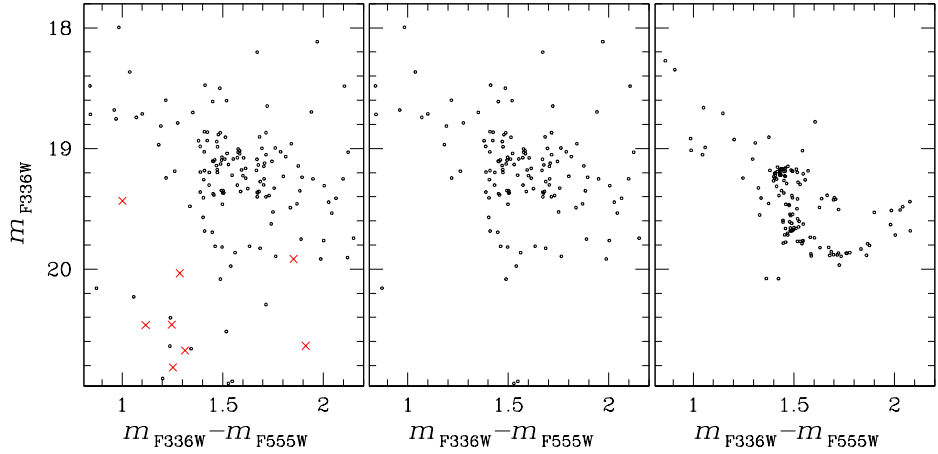} 
%/home/milone/MC/NGC1856/all/MATCH/figure/fig.macro go4d
 \caption{ \textit{Left panel:} $m_{\rm F336W}$ vs.\,$m_{\rm F336W}-m_{\rm F555W}$ CMD zoomed around the red clump. Black circles and red crosses are stars in the cluster and reference field, respectively. \textit{Middle panel:}  CMD of stars in the cluster field after that reference-field stars have been statistically subtracted. \textit{Right panel:} Simulated CMD of a 300\,Myr-old simple stellar population (see text for details).}
 \label{fig:RC} 
\end{figure*} 
\end{centering}

  Left panel of Fig.~\ref{fig:RC} shows a zoom of the $m_{\rm F336W}$ vs.\,$m_{\rm F336W}-m_{\rm F555W}$ around the red clump. The 142 black points represent stars in the cluster field, while the 8 reference-field stars are plotted with red crosses. The contamination from field stars is lower than $\sim$6\%. We have used the same procedure of Sect.~\ref{sub:field} to statistically subtract field stars from the cluster-field CMD and plot the CMD of subtracted stars in the middle panel of Fig.~\ref{fig:RC}.  
 The red clump is widely spread in color and magnitude and spans a range of $\sim$0.4 mag in $m_{\rm F336W}-m_{\rm F555W}$ and show same hint of bimodality, with two main components at ($m_{\rm F336W}-m_{\rm F555W}$:$m_{\rm F336W}$)=(1.50:19.25) and (1.60:19.25).
 We have used solar-scaled BaSTI isochrones with Z=0.01, and [Fe/H]=$-$0.25,  to simulate a simple stellar population with t=300\,Myr. We assumed the same errors as in the observed CMD, and the fraction of binaries measured in Sect.~\ref{binarie}.  The simulated CMD is plotted in the right panel of Fig.~\ref{fig:RC}, where we plotted the same number of stars as in the middle-panel CMD. A visual comparison of the observed and the simulated CMD immediately reveals that the red clump of NGC\,1856 is not consistent with a simple population.

In principle, the red clump could be used to constrain the star-formation history of NGC\,1856. However, the luminosity and the color of the red clump depends not only on age but also, on several parameters like the amount of core overshooting, the mass loss,  and the helium content (see e.g.\,Girardi et al.\,2009). Unfortunately, these quantities are not constrained for NGC\,1856, and we prefer to not use the red clump to investigate the star-formation history of NGC\,1856.  

%%%%%%%%%%%%%%%%%%%%%%%%%%%%%%%%%%%%%% FIG 3 %%%%%%%%%%%%%%%%%%%%%%%%%%%%%%%%%%% 
%%%%%%%%%%%%%%%%%%%%%%%%%%%%%%%%%%%%%%%%%%%%%%%%%%%%%%%%%%%%%%%%%%%%%%%%%%%%%%% 

\section{Summary and Discussion} 
We have used multi-band UVIS/WFC3 photometry to study the LMC star cluster NGC\,1856. We found that the MSTO is broadened, and that, below the turn-off, the MS splits into two main components. In the magnitude interval 19.9$<m_{\rm F555W}<$20.6 the blue and the red MSs host 33$\pm$5\% and 67$\pm$5\% of the total number of MS stars, respectively. 

We have demonstrated that both the split MS and the eMSTO can not be due to photometric errors, differential reddening, or unresolved binaries. 
 As discussed in Sect.~\ref{sec:intro}, the eMSTO is a common feature of intermediate-age star clusters in the Large and the Small Magellanic Cloud although its interpretation is still controversial. 
This paper shows the first evidence of an eMSTO in a young star cluster with age less than $\sim$1 Gyr and provides new steps towards the understanding of this phenomenon. 
We have compared the observed CMD with isochrones from the BaSTI archive and estimated an average age of $\sim$300 Myr. 
 We have demonstrated that,  if stars in NGC\,1856 are chemically homogeneous and the eMSTO is due to a prolonged star formation,
the latter should have lasted $\sim$150 Myr, not necessarily in a continuous way, in view of the features we observe in the upper MS and MSTO. 
Specifically, the interval 
between the 90$^{\rm th}$ and the 10$^{\rm th}$  percentile of the age distribution corresponds to
$\sigma_{80}=174\pm$31 Myr. The age difference between the 84$^{\rm th}$ and the 16$^{\rm th}$ percentile is  $\sigma_{68}=139\pm$22 Myr.
In the hypothesis that the MS and MSTO broadening are due to a prolonged star formation, the red MS evolves into both the faint and the bright MSTO, and the blue MS into the bright MSTO.
 
The eMSTO of intermediate-age MC clusters have been interpreted either with an age spread or with stellar rotation. The idea that this feature of the CMD is due to prolonged star formation has been suggested by several authors (e.g.\,Mackey et al.\,2008; Glatt et al.\,2008; Paper\,I; Goudfrooij et al.\,2011) and has been recently supported by Goudfrooij et al.\,(2014) on the basis of their study of 18 MC clusters. These authors have found a correlation between the MSTO width and the escape velocity ($v_{\rm esc}$) of the host cluster and argued that these results are unlikely due to stellar rotation. 
  
 Goudfrooij et al.\,(2014)  and Correnti et al.\,(2014) noticed that all the analyzed clusters have $v_{\rm esc} \geq$ 17 km s$^{-1}$ and suggested that eMSTO can occur only in clusters whose escape velocities are higher than the winds of first-generation stars from which the second generations formed. Intermediate-mass asymptotic-giant-branch stars and massive binaries have wind velocities close to 12-15\,km s$^{-1}$, thus being possible candidates for producing the material from which the second generation formed (see Goudfrooij et al.\,2014  and Correnti et al.\,2014 for details).  Moreover, Goudfrooij,  Correnti and collaborators have hypothesized the existence of an early-escape-velocity threshold of about 15 km s$^{-1}$ which differentiate clusters that have experienced a prolonged star-formation from those without eMSTO.   
 
 The evidence of an eMSTO in NGC\,1856, whose velocity escape did not exceed $\sim$15\,km s$^{-1}$ at its formation  (Goudfrooij et al.\,2014), is consistent with this picture.    
 In this context, it would be interesting to understand why the $\sim$180-Myr old LMC cluster NGC\,1866, whose velocity escape is only slightly smaller than that of NGC\,1856 does not show an eMSTO (Bastian \& Silva-Villa\,2013).  

 We have investigated the possibility that the double MS and the eMSTO have different metallicity and C$+$N$+$O abundance. In these cases the blue MS would evolve into the faint MSTO, while bright-MSTO stars would be the progeny of the red MS. 
 However, also in case of iron variation among stars in NGC\,1856, we need to assume
an age difference of $\sim$150 Myr.
If we assume that NGC\,1856 has constant overall CNO content, the blue MS and the faint MSTO correspond to a $\sim$400\,Myr population with [Fe/H]=$-$0.6, while the rMS and the bMSTO are consistent with a $\sim$150\,Myr-younger stellar population enhanced in [Fe/H] by $\sim$0.4 dex.  Observations are also consistent with the younger population having almost the same iron content as the old ones but being enhanced in C$+$N$+$O by about a factor of two. 

 Interestingly, in Paper\,II we have found that the $\sim$150-old LMC cluster NGC\,1844 exhibits an intrinsic color spread along the MS. We have shown that such a feature is well fitted by isochrones with different C$+$N$+$O, although these models provide a poor fit of the bright MS of NGC\,1844. Due to the possible similarity between the MS of NGC\,1844 and NGC\,1856 it is tempting to speculate that C$+$N$+$O variations could play a role on the morphology of the MS and MSTO morphology of young and intermediate-age MC clusters.

Unfortunately no abundance measurements are available for
NGC\,1856 to date. Moreover, we are not able to establish whether NGC\,1856 has homogeneous C, N, O and Fe from photometry alone because the connection between the faint and the bright MSTO with the two MSs can not be established unequivocally from our CMDs.
In this context, it is worth noting the finding by Mucciarelli et al.\,(2014) that in the eMSTO of the LMC cluster NGC\,1846 there is no evidence for significant variation of [O/Fe], [Na/Fe], [Al/Fe], [Mg/Fe], and [Fe/H]. This result suggests that intermediate-age clusters in MCs could have homogeneous iron and light-element abundance. A chemical-abundance study of NGC\,1856 is needed in order to estimate the iron and C, N, O content of this cluster.
  
As an alternative interpretation of the eMSTO phenomenon, Bastian \& De Mink\,(2009) have suggested that rotation in stars mimic an age spread and can be responsible for the eMSTO of intermediate-age MC clusters. This idea is supported by the fact that the structure of a star is significantly affected by rotation and that  its effective temperature depends on the inclination angle relative to the observer.  
  This interpretation however is in disagreement with the results by Girardi et al.\,(2011), who have calculated isochrones for intermediate-age MC clusters and concluded that a dispersion of radial velocity can not be the only reason for the eMSTO. 

Platais et al.\,(2012) have measured projected rotational velocities for upper-MS stars of the $\sim$1.3 Gyr old Galactic open cluster Trumpler\,20 and found very-high rotations up to $v$ sin{\it i}$\sim$180 km\,s$-1$. These authors have selected two groups of slow- and fast-rotators and found that fast rotators have a  marginal bluer colors than slow rotators ($\delta$($V-I$)$\sim -$0.01 mag). 
 They concluded that the MSTO morphology of Trupler\,20 is marginally affected by rotation. Unfortunately no direct measurements of rotational velocities are available for MC clusters with eMSTO.

An accurate comparison of our observations with appropriate stellar models is certainly needed to understand the importance of stellar rotation for the explanation of the eMSTO in NGC 1856. Nevertheless our findings provide a major challenge to the scenario suggested by Bastian \& De Mink.  
Indeed, in order to explain the eMSTO of intermediate-age clusters, these authors have suggested  that only stars with masses $1.2<\mathcal{M}< 1.7 \mathcal{M}_{\odot}$ would be affected by rotation. 
 In the young NGC\,1856, stars in this mass interval are located on the MS region, with 20.7$<m_{\rm F336W}<$23.4, where there is no evidence for a large spread in color, thus suggesting that stellar rotation, if present, would not significantly affect the colors and the magnitudes of these stars.  
If the eMSTO of both NGC\,1856 and intermediate-age clusters is due to stellar rotation it would be challenging to understand why this phenomenon would affect stars with different masses in clusters with different ages.  

Yang et al.\,(2011) have suggested merged binary systems and interactive binaries with mass transfer can be responsible of both the eMSTO and dual clumps in intermediate-age MC clusters. 
 Despite this the rest of the MS of these clusters is not significantly spread by binary interaction.  
 The evidence that NGC\,1856 exhibits a dual MS seems in disagreement with the scenario suggested by Yang and collaborators. However, it should be emphasized that conclusions from these authors are based on intermediate-age star clusters and that appropriate studies of young clusters are mandatory to establish if interacting binaries are consistent with observations of NGC\,1856 or not. 
 
In summary the detection of a double MS and an eMSTO in the young NGC\,1856 adds new information to the puzzling eMSTO-phenomenon of MC star clusters. 
 On the one hand, the new findings seem consistent with a prolonged star-formation history thus challenging the interpretation that either stellar rotation or interacting binaries can be responsible of the eMSTO.
  On the other hand, the recent findings that the SGB morphology of some massive intermediate-age MC cluster with eMSTO  is not consistent with a spread in age (Bastian \& Niederhofer\,2015; Li et al.\,2014), and the lack of evidence of prolonged star-formation history in extragalactic massive clusters  raise doubts on the interpretation of eMSTO as due to multiple  stellar generations with different ages (or prolonged star formation).

\section*{acknowledgments} 
\small 
 We thank the referee for a constructive report which has improved the quality of the manuscript.
We warmly thank Jay Anderson who has provided most of the programs for the data reduction and Aaron Dotter for the unpublished values of the extinction rate of  the UVIS/WFC3 filters. SC and GP acknowledge partial support by PRIN-INAF 2014. SC acknowledges partial support by PRIN MIUR 2010-2011, project \lq{The Chemical and Dynamical Evolution of the Milky Way and Local Group Galaxies}\rq, prot. 2010LY5N2T.  APM and HJ acknowledge support by the Australian Research Council through Discovery Early Career Researcher Award DE150101816 and Discovery Project DP150100862.

\bibliographystyle{aa}

\end{document}